\documentclass[longauth]{aa}

\usepackage{graphicx}

\usepackage{xcolor}

\usepackage[normalem]{ulem}

\usepackage{txfonts}

\usepackage{hyperref}

\begin{document} 

\title{CoSEE-Cat: A Comprehensive Solar Energetic Electron event Catalogue obtained from combined in situ and remote-sensing observations from Solar Orbiter}
\subtitle{Catalogue description and first statistical results}
\titlerunning{CoSEE-Cat: A Comprehensive Solar Energetic Electron event Catalogue obtained from Solar Orbiter}

   \author{A.\,Warmuth \inst{1}
          \and
          F.\,Schuller \inst{1}
          \and
          R.\,G\'omez-Herrero\inst{2}
          \and
          I.\,Cernuda \inst{2}
          \and
          F.\,Carcaboso \inst{3, 4, 5}
          \and
          G.\,M.\,Mason \inst{6}
          \and
          N.\,Dresing \inst{7}
           \and
          D.\,Pacheco \inst{8}
          \and
          L.\,Rodríguez-García \inst{9,2}
          \and
          M.\,Jarry \inst{10,11}
          \and
          M.\,Kretzschmar \inst{12}
          \and 
          K.\,Barczynski \inst{13, 14}
          \and
          D.\,Shukhobodskaia \inst{15}
          \and
          L.\,Rodriguez \inst{15}
          \and
          S.\,Tan \inst{1}
           \and
          D.\,Paipa-Leon \inst{16}
          \and
          N.\,Vilmer \inst{16}
          \and
          A.\,P.\,Rouillard\inst{10}
          \and
          C.\,Sasso \inst{17}
          \and
          S.\,Giordano \inst{18}
          \and
          G.\,Russano \inst{17}
          \and
          C.\,Grimani \inst{19, 20}
          \and
          F.\,Landini \inst{18}
          \and
          C.\,Mac Cormack \inst{21, 4}
          \and
          J.\,A.\,J.\,Mitchell \inst{1}
          \and
          A.\,Fedeli \inst{7}
          \and
          L.\,Vuorinen \inst{7, 22}
          \and
          D.\,Lario \inst{4}
          \and
          H.\,A.\,S.\,Reid \inst{23}
          \and
          F.\,Effenberger \inst{24}
          \and
          S.\,Musset \inst{6}
         \and
          K.\,Riebe \inst{1}
          \and
          A.\,Galkin \inst{1}
          \and
          K.\,Makan \inst{1}
          \and
          S.\,Reusch \inst{1}
          \and
          A.\,Vecchio \inst{25,15}
          \and
          O.\,Dudnik \inst{26,27}
          \and
          S.\,Krucker \inst{28}
          \and
          M.\,Maksimovic \inst{16}
          \and 
          J.\,Rodr\'iguez-Pacheco\inst{2}
          \and
          M.\,Romoli \inst{29, 30}
          \and
          R.\,F.\,Wimmer-Schweingruber \inst{31}
          }

   \institute{Leibniz-Institut f\"ur Astrophysik Potsdam (AIP), An der Sternwarte 16, 14482 Potsdam, Germany
              \email{awarmuth@aip.de}
    \and
    Universidad de Alcal\'a, Space Research Group, 28805 Alcal\'a de Henares, Spain
    \and
    Postdoctoral Program Fellow, NASA Goddard Space Flight Center, Greenbelt, MD, USA
     \and
     Heliospheric Physics Laboratory, Heliophysics Science Division, NASA Goddard Space Flight Center, 8800 Greenbelt Rd., Greenbelt, MD 20770, USA
     \and
     Goddard Planetary Heliophysics Institute, University of Maryland, Baltimore County, Baltimore, MD, 21250, USA
     \and
     Applied Physics Laboratory, Johns Hopkins University, Laurel, MD, 20723, USA
    \and
    Department of Physics and Astronomy, 20014 University of Turku, Finland
     \and
     Deep Space Exploration Laboratory/School of Earth and Space Sciences, University of Science and Technology of China, Hefei 230026, China
    \and European Space Agency (ESA), European Space Astronomy Centre (ESAC), Camino Bajo del Castillo s/n, 28692 Villanueva de la Cañada, Madrid, Spain
    \and
    Institut de Recherche en Astrophysique et Planétologie (IRAP), CNRS, Université de Toulouse III-Paul Sabatier, Toulouse, France
    \and
    Institute for Astronomy, Astrophysics, Space Applications and Remote Sensing (IAASARS), National Observatory of Athens (NOA), Penteli, Greece
     \and
     LPC2E UMR7328, OSUC/Université d’Orléans/CNRS/CNES, 3a av de la recherche scientifique, 45071 Orléans, France
     \and
     Physikalisch-Meteorologische Observatorium (PMOD/WRC), Dorfstrasse 33, 7260 Davos Dorf, Switzerland
     \and
     ETH-Zurich, Hönggerberg Campus, HIT Building, Wolfgang-Pauli-Str. 27, 8093 Zürich, Switzerland
    \and
     Solar-Terrestrial Centre of Excellence – SIDC, Royal Observatory of Belgium; Avenue Circulaire 3, 1180 Brussels, Belgium
     \and
     LIRA, Observatoire de Paris, PSL Research University, CNRS, Sorbonne Université, Université Paris Cité, 5 place Jules Janssen, 92195 Meudon, France
     \and
     INAF - Astronomical Observatory of Capodimonte, Salita Moiariello 16, I-80131 Napoli, Italy
     \and
     INAF - Astrophysical Observatory of Torino, Via Osservatorio 20, I-10025 Pino Torinese, Italy
     \and
     University of Urbino Carlo Bo, Department of Pure and Applied Sciences, Via Santa Chiara 27, I-61029 Urbino, Italy
     \and
     INFN, Section in Florence, Via Bruno Rossi 1, I-50019 Florence, Italy
     \and
     The Catholic University of America, Washington, DC 20064, USA
     \and
     Department of Physics and Astronomy, Queen Mary University of London, Mile End Road, London E1 4NS, UK
     \and
     Mullard Space Science Laboratory, University College London, Holmbury St. Mary, Dorking, Surrey, RH5 6NT, UK
    \and
    Ruhr University Bochum, Bochum, Germany
    \and 
    Radboud Radio Lab, Department of Astrophysics, Radboud University, Nijmegen, The Netherlands
    \and
    Space Research Center of Polish Academy of Sciences, Warsaw, Bartycka str., 18A, 00-716 Poland
    \and
    Institute of Radio Astronomy of National Academy of Sciences of Ukraine, Kharkiv, Mystetstv str., 4, 61002 Ukraine
    \and
    University of Applied Sciences and Arts Northwestern Switzerland (FHNW), Bahnhofstrasse 6, 5210, Windisch, Switzerland
     \and
     University of Florence, Department of Physics and Astronomy, Via Giovanni Sansone 1, I-50019 Sesto Fiorentino, Italy
     \and
     INAF-Astrophysical Observatory of Arcetri, Largo Enrico Fermi 5, I-50125 Firenze, Italy
     \and
     Institute of Experimental and Applied Physics, Kiel University, D-24118 Kiel, Germany
}

   \date{}

  \abstract
   {The acceleration of particles at the Sun and their propagation through interplanetary space are key topics in heliophysics. Specifically, solar energetic electrons (SEEs) measured in situ can be linked to solar events such as flares and coronal mass ejections (CMEs) since they are also observed remotely in a broad range of electromagnetic emissions such as in radio and X-rays. Solar Orbiter, equipped with a wide range of remote-sensing and in situ detectors, provides an excellent opportunity to investigate SEEs and their solar origin from the inner heliosphere.}
   {We aim to record all SEE events measured in situ by Solar Orbiter, and to identify and characterise their potential solar counterparts. The results have been compiled in the \textit{Comprehensive Solar Energetic Electron event Catalogue} (CoSEE-Cat), which will be updated regularly as the mission progresses. The catalogue contains key parameters of the SEEs, as well as the associated flares, CMEs, and radio bursts. In this paper, we describe the catalogue and provide a first statistical analysis.}
   {The Energetic Particle Detector (EPD) was used to identify and characterise SEE events, infer the electron release time at the Sun, and determine the composition of related energetic ions. Basic parameters of associated X-ray flares (timing, intensity, source location) were provided by the Spectrometer/Telescope for Imaging X-rays (STIX). This was complemented by the Extreme Ultraviolet Imager (EUI), which added information on eruptive phenomena. CME observations were contributed by the coronagraph Metis and the Solar Orbiter Heliospheric Imager (SoloHI). Type III radio bursts observed by the Radio and Plasma Waves (RPW) instrument provided a link between the SEEs detected at Solar Orbiter and their potential solar sources. The conditions in interplanetary space were characterised using Solar Wind Analyzer (SWA) and Solar Orbiter Magnetometer (MAG) measurements. Finally, data-driven modelling with the Magnetic Connectivity Tool provided an independent estimate of the solar source position of the SEEs.}
   {The first data release of the catalogue contains 303 SEE events observed in the period from November 2020 until the end of December 2022. Based on the timing and magnetic connectivity of their solar counterparts, we find a very clear distinction between events with an impulsive ion composition and ones with a gradual one. These results support the flare-related origin of impulsive events and the association of gradual events with extended structures such as CME-driven shocks or erupting flux ropes. We also show that the commonly observed delays of the solar release times of the SEEs relative to the associated X-ray flares and type III radio burst are at least partially due to propagation effects and not exclusively due to an actual delayed injection. This effect is cumulative with heliocentric distance and is probably related to turbulence and cross-field transport.}
   {}


   \maketitle

\section{Introduction}
\label{sec:intro}

The Sun is the most energetic particle accelerator in the solar system. Ions and electrons accelerated at or near the Sun can escape into interplanetary (IP) space where they are detected in situ as solar energetic particles \citep[SEPs; e.g.][]{Reames1999}. Accelerated particles can also be guided by coronal magnetic field lines to lower layers of the solar atmosphere, where they interact with the ambient medium, which dissipates their energy, heats plasma, and generates non-thermal emission in X-rays and $\gamma$-rays that can be observed with remote-sensing instruments \citep[e.g.][]{Fletcher2011,Holman2011,Vilmer2011,Warmuth2020}.  

Solar energetic particle events generally fall into two broad classes. Gradual events are associated with large X-ray flares and coronal mass ejection (CME)-driven shocks, and can be measured over wide heliolongitudinal spans with respect to the parent solar eruption. Impulsive events are electron-rich and associated with small X-ray flares, type III radio bursts, and highly enriched ion abundances in $^3$He \citep[e.g. reviews by][]{Desai2016,Reames2018}. Although these terms originated from the time evolution of the associated soft X-ray (SXR) flares \citep{Cane1986}, they are now commonly used to indicate the elemental composition of SEPs \citep{Reames1999}. 

Solar energetic electron (SEE) events are electron intensity enhancements detected in IP space at energies from a few kilo-electronvolts to a few mega-electronvolts. They are usually also accompanied by ion enhancements and can therefore occur in association with impulsive and gradual SEP events. Impulsive SEE events are highly associated with X-ray flares and type III radio bursts \citep{Lin1985,Wang2012}. In these events, the characteristics of the precipitating energetic electrons (i.e. them being downward-moving) can be inferred from hard X-ray (HXR) observations, while type III radio emission allows escaping energetic electrons to be traced through the corona into IP space \citep{Kane1981,Hamilton1990,Reid2017}. The availability of these complementary observations implies that SEEs provide an ideal opportunity to study particle acceleration and transport in an astrophysical system.

The flare-related origin of impulsive SEEs is supported by several lines of evidence, including temporal associations of the inferred solar release times (SRTs; i.e. the times at which the particles are injected into IP space) of SEEs with HXR flares and type III radio bursts, and correlations between the number and spectral index of precipitating and escaping electrons \citep{Krucker2007,Dresing2021}. However, there are some long-standing inconsistencies that raise questions about the interpretation that the upward- and downward-moving electron populations in flares are really accelerated by the same mechanism. Although there are ‘prompt’ SEE events that appear to be injected at the peak of the associated HXR flare and type III burst \citep[e.g.][]{Krucker2007}, most of the SEE events show apparent SRT delays. The average delay is of the order of 10~mins \citep{Haggerty2002}. Another issue concerns the relation between the spectral indices of the SEEs and the electrons in the associated flare. When energetic electron properties are inferred from observed HXR spectra, either a thick-target or a thin-target bremsstrahlung model is assumed. Although the SEE spectra and the inferred flare electron spectra do indeed correlate, the results are not consistent with either model \citep{Krucker2007,Dresing2021}.

Two scenarios could explain the discrepancies in timing and spectra: either there are intrinsic differences in the acceleration and/or injection of the flare electrons and the SEEs \citep{Wang2006}, or alternatively the upward- and downward-moving electron populations are initially similar, but transport effects in the IP medium then modify their characteristics, including wave-particle interactions \citep{Vocks2012,Reid2013,Reid2017b} and pitch-angle scattering \citep{Droege2003,Droege2018}. A combination of acceleration differences and transport effects is also possible.  However, so far it has not been possible to disentangle acceleration and/or injection from transport effects.

One approach to addressing this question is to sample SEEs at different heliocentric distances. If transport effects influence the electrons travelling through IP space, a systematic change in the relation between SEE and flare electron properties is expected with varying heliocentric distance. The Solar Orbiter mission \citep{Muller2020} provides us with an excellent opportunity to investigate SEEs and their solar origin from inner heliospheric distances. Due to its elliptical orbit, the mission samples heliocentric distances from 0.28 au to approximately 1 au. Additionally, Solar Orbiter is equipped with all required in situ and remote-sensing instruments on a single platform, providing a comprehensive dataset and high duty cycle required to analyse SEE events \citep{GomezHerrero2021,Lorfing2023}

Our goal is to document all SEE events measured in situ by Solar Orbiter and to identify and characterise their potential solar counterparts. These tasks have been carried out by a joint working group involving team members from eight of the Solar Orbiter instruments: EPD, STIX, EUI, RPW, Metis, SoloHI, SWA, and MAG. We include basic information on these instruments in the following section (Sect. \ref{sec:data}). All information derived by the team is included in CoSEE-Cat, which can be accessed online\footnote{\url{https://coseecat.aip.de/}\label{cosee_cat}}. The first data release of CoSEE-Cat includes SEE events from November 2020 up to the end of 2022. However, CoSEE-Cat is a living catalogue that will be updated as the Solar Orbiter mission progresses. 

In this paper, we present CoSEE-Cat, including an overview of the instruments and datasets used (Sect.~\ref{sec:data}), and detailed explanations of how the SEE events were identified and the parameters derived (Sect.~\ref{sec:catalog}). In addition, we report the results of a first statistical study of the SEE events (Sect.~\ref{sec:results}). The conclusion of this study is given in Sect.~\ref{sec:conclusion}. The specific parameters provided by the event catalogue are listed in Appendix~\ref{sec:contents}, and a description of additional resources available in the online version of the catalogue is provided in Appendix~\ref{sec:online}.


\section{Data and instrumentation}
\label{sec:data}

This study combines observational data from a large number of Solar Orbiter's in situ and remote-sensing instruments, supplemented by data-driven modelling of the magnetic connectivity of Solar Orbiter to the Sun. We used the Energetic Particle Detector (EPD) suite of instruments \citep[][]{Rodriguez-Pacheco2020, Wimmer2021} to identify and characterise SEE events. More specifically, we used the instrument units SupraThermal Electrons and Protons (STEP), Electron Proton Telescope (EPT), and High Energy Telescope (HET) instrument units, which cover electron energies of 2--80~keV, 25--475~keV, and 0.3--30~MeV, respectively. In addition, we used the Suprathermal Ion Spectrograph (SIS) unit to determine the composition of the associated energetic ions in the energy range of $\sim 0.1-10$~MeV per nucleon.

Solar flares associated with these electron events were studied primarily with the Spectrometer/Telescope for Imaging X-rays \citep[STIX;][]{Krucker2020}. STIX provides imaging spectroscopy in the X-ray range from 4 to 150~keV and consequently constrains both the hot plasma and the accelerated electrons in flares via remote sensing. It has a full-disc field of view (FOV) and sub-second time resolution. Most relevant to this study, STIX provided quantitative information on the timing, location, intensity, and spectra of the energetic electrons. 

The Extreme Ultraviolet Imager \citep[EUI;][]{Rochus2020}  was used to provide additional context information on associated flares and eruptive phenomena. It comprises three instruments: a Full Sun Imager (FSI), providing observations in 174 and 304 \AA\ and two High Resolution Imagers (HRIs), imaging in the Lyman-alpha line of hydrogen at 1\,216 \AA\ and at 174 \AA. FSI has an unprecedentedly large FOV, which at perihelion covers $4~R_\odot$, ensuring that the full solar disc is always visible. At 1~au the FOV reaches 14.3~$R_\odot$,  providing the possibility to trace eruptive material through this region. The cadence of FSI images depends on the observing mode, ranging from 2 minutes to 1 hour, with most images taken at 10-minute intervals. Data used in this study are from EUI data release 6.0 2023-01\footnote{\url{https://doi.org/10.24414/z818-4163}}. 

The Radio and Plasma Waves \citep[RPW;][]{Maksimovic2020} instrument was used to identify and characterise type~III radio bursts, which provide a link between the in situ particle events and their solar sources. Specifically, we used electric fields measurements provided by the Thermal Noise Receiver (TNR) and High Frequency Receiver (HFR), which cover the frequency range from 4~kHz up to 16~MHz.

The multi-channel imaging coronagraph Metis \citep{Antonucci2020} is capable of simultaneously observing the solar corona in the ultraviolet (UV) narrow band centered around the  1\,216~\AA\ Ly-$\alpha$ line emitted by neutral H atoms, in addition to the classical visible-light (VL) polarised broadband in the interval 580--640~nm. The instrument FOV is an annulus extending from 1.6$^\circ$ to 2.9$^\circ$ radius.
Metis can therefore cover projected altitude intervals going from $1.7-3.1~R_\odot$ (when Solar Orbiter is at 0.28 au, minimum perihelion) to $2.8-5.5~R_\odot$ (at 0.5~au), up to $6.0-12.9~R_\odot$ (at 1.02~au, maximum aphelion), in the course of the eccentric orbit of Solar Orbiter. The instrument plate scale is 10.7"~pixel$^{-1}$ and 20"~pixel$^{-1}$ in the VL and UV channels, respectively \citep[][]{Fineschi2020}.

The Solar Orbiter Heliospheric Imager \citep[SoloHI;][]{Howard2020} provides white-light images of the inner heliosphere. The imaging plane of the SoloHI telescope is made up of four tiles that conform to a FOV of $40^{\circ}$ starting at $5^{\circ}$ off the east limb of the Sun relative to the Solar Orbiter. At perihelion, SoloHI presents an effective resolution comparable to the C2 telescope of the coronagraph Large Angle and Spectrometric Coronagraph Experiment \citep[LASCO]{brueckner_1995} on board the Solar and Heliospheric Observatory \citep[SOHO]{domingo_1995} mission, while offering a wider FOV (6–60$\,$R$_{\odot}$) and a higher signal-to-noise ratio than the SOHO/LASCO-C3 telescope, the FOV of which extends up to 32$\,$R$_{\odot}$.

Finally, we characterised the conditions of the IP medium in which the energetic electrons propagate with two additional in situ instruments on board Solar Orbiter. The Proton and Alphas Sensor (PAS), part of the  Solar Wind Analyzer \citep[SWA;][]{Owen2020} suite, was used to provide solar wind parameters, including density, temperature, and bulk flow speed, while the Solar Orbiter Magnetometer \citep[MAG;][]{Horbury2020} was used to measure the IP magnetic field (IMF) vector.

For events associated with flares visible from Earth, the data from Solar Orbiter were supplemented by the SXR fluxes routinely provided by the X-ray Sensor (XRS) aboard the Geostationary Operational Environmental Satellite (GOES).


\section{Event selection and parameter determination}
\label{sec:catalog}

\subsection{Energetic electron events}
\label{sec:see}

\begin{figure}
    \centering
     \includegraphics[width=\linewidth]{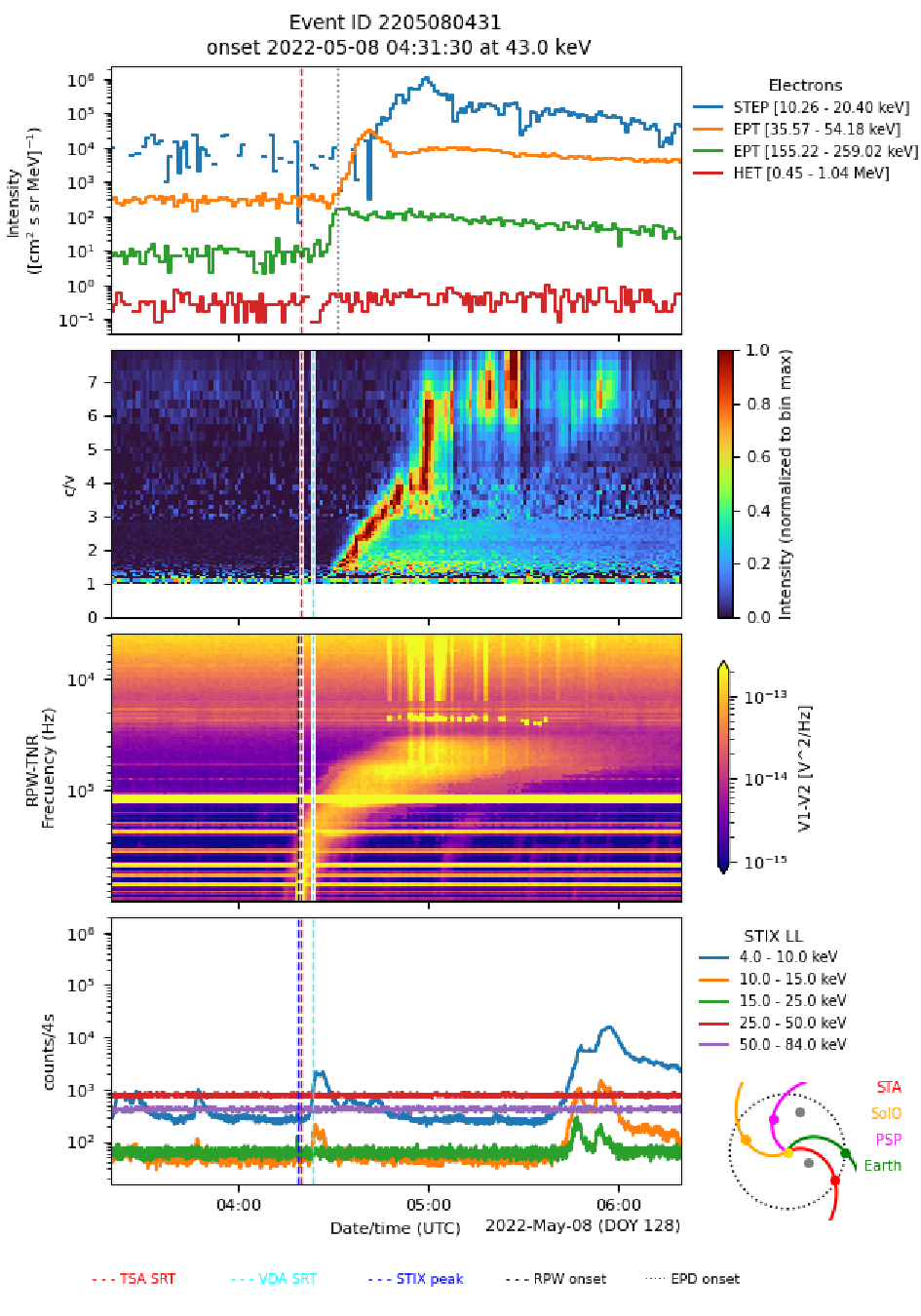}
    \caption{EPD, RPW, and STIX combined observations during an SEE event observed on 2022 May 8 (event ID \textit{2205080431}). Top panel: Omnidirectional electron time profiles observed by different EPD sensors at selected energy bands between 10.26 keV and 1.04 MeV. Second panel: c/v versus time plot for EPD electrons. The electron intensities were normalised to the maximum flux observed in each energy band,  as is indicated by the colour bar, for a clearer visualisation of the velocity dispersion. Third panel: RPW dynamic radiospectrum, showing a type III radio burst close to the electron release time. Bottom panel: X-ray light curves observed by STIX in five different energy bands from 4 to 84 keV. The small insert at the bottom right shows the locations of STEREO-A, Solar Orbiter, PSP and the inner planets. The times indicated by the vertical lines are defined at the bottom of the plot.}
    \label{fig:epd_stix_rpw_example}
\end{figure}

In this paper, we identified and analysed SEE events observed by EPD from November 2020 to the end of 2022. We used the level 2 EPD datasets publicly available in the Solar Orbiter Archive (SOAR\footnote{\url{https://soar.esac.esa.int/}}). The initial step in the selection of the events involved inspecting 1-minute averaged electron fluxes above 10 keV observed by STEP, EPT, and HET and searching for periods of enhanced electron intensity exceeding the sensors' background levels. Only intensity enhancements showing at least three consecutive points above the pre-event background plus three standard deviations in at least one EPD energy channel were considered. 
Figure \ref{fig:epd_stix_rpw_example} shows an SEE event included in the catalogue, displaying the increased electron fluxes at different EPD channels in the top panel as indicated in the legend, along with the identification of the onset time at 44 keV (dotted line). Note that this onset time is also encoded in the event ID which is used as a unique identifier for all SEE events in the catalogue. Thus the EPD onset on 2022~May~8 at 04:31~UT corresponds to event ID \textit{2205080431}. The second panel shows the electron's inverse velocity, c/v, with the colour map normalised to the maximum flux in each energy band to help the visualisation of the event. The third and fourth panels show the RPW dynamic radiospectrum and the X-ray light curves observed by STIX as described in sections \ref{sec:flare_stix} and \ref{sec:type3}. Finally we included some orbital context information at the bottom right with the locations of STEREO-A, Solar Orbiter, Parker Solar Probe, and the inner planets.

An event was then selected if it indicated a solar rather than a solar wind or planetary origin. The presence of velocity dispersion, shown in the second panel of Fig. \ref{fig:epd_stix_rpw_example}, was taken as clear evidence of a solar origin.  However, the presence of this velocity dispersion was not considered a requirement in the selection of the events, as it may have been missing or unclear for various reasons (e.g. low statistics, interrupted or intermittent magnetic connection, transport effects, a narrow energy interval, etc.). We included only events with clear timing information and excluded periods with a large number of overlapping electron intensity enhancements within a short time interval that could not be resolved. This selection resulted in a total of 303 SEE events detected at heliocentric distances ranging from 0.3 au to 1.0 au.

The SEE event identification was normally based on EPT electron fluxes in the reference energy band of $35.6-54.2$ keV (geometric mean of 44~keV) of the sunward-looking telescope. It should be noted that the reference value has changed slightly between different EPT calibrations from 43~keV to the current 44~keV. Lower energy channels from STEP were used when the SEE event was not clearly visible in the EPT data. The different FOVs of the EPT instrument\citep{Rodriguez-Pacheco2020} were used as appropriate for each observation.
We used a time averaging of 1 minute as a baseline, but for a subset of short or rapidly rising events, a higher cadence was used when the counting statistics were sufficiently high. Conversely, some events with extremely slow rise profiles and/or poor statistics required longer averages.

Each SEE event in the dataset is characterised by parameters measured by EPD or derived from its measurements, such as the electron onset time and peak intensity at the reference energy, the particle's SRT
determined by the time shift analysis (TSA) and velocity dispersion analysis (VDA) methods, as is discussed in Sect. \ref{sec:epdtiming}. The peak time, which indicates the time of maximum intensity of the prompt component of the SEE event \citep[e.g.][]{Lario2013, 2023Rodriguez-Garcia} at the reference energy, is also provided.

Each SEE event was also analysed for its elemental and isotopic composition of ions H--Fe measured by SIS and classified by the degree of anisotropy using the absolute value of the first-order anisotropy (small, medium, large), as detailed in Sect.~\ref{subsec:anisotropy} and Sect.~\ref{subsec:composition}.

\subsubsection{EPD solar release times}\label{sec:epdtiming}

 Energetic particle timing information is a key parameter needed to link SEE observations to solar events. We estimated the SRT of the energetic electrons, i.e. the time when the electrons are injected at the Sun, using TSA, and, whenever possible, using VDA too, which are the techniques frequently used for this purpose \citep[e.g.][]{Vainio2013}. TSA assumes that the first-arriving particles propagate scatter-free with no energy loss along an ideal Parker spiral and uses a single energy channel to infer the time of particle release at the Sun. 

 The electrons' SRT, $t_{srt}$, for a given energy, $E$, was obtained by time-shifting the particle's onset time at the spacecraft by the travel time of the particles along the IP magnetic field \citep{Vainio2013, Paassilta2018}:
\begin{equation}
t_{srt}(E) = t_{o}(E) - \frac{L_n(v_{sw})}{v(E)}
\label{eq:tsa}
\end{equation} 
where the onset time $t_{o}$ is the first time the electron flux exceeds the 3-sigma level above the background level and continues enhanced at least three consecutive points at the reference energy. $L_n$ is the nominal Parker spiral length, and $v(E)$ is the electron kinetic speed according to the employed energy $E$. We used the solar wind speed observed by SWA on board Solar Orbiter at the time of SEE onset to calculate $L_n$ from the Parker spiral model assumed to be valid from the spacecraft to the solar surface. When no SWA measurements were available, a nominal value, $v_{sw} = 400$\,$ \mathrm{km/s}$, was assumed. 

The VDA technique \citep{Reames1985,Vainio2013} assumes a simultaneous release from the Sun of all electrons of different energies, followed by scatter-free propagation for the first-arriving particles with no energy loss following a single effective path length, $L$. In this case $t_{srt}$ and $L$ are free parameters obtained by a linear fit of a series of onset times versus inverse particle speeds:  
\begin{equation}
    t_{o}(E) = t_{srt} + \frac{L}{v(E)}
,\end{equation} where $t_{o}(E)$ are the onset times and $v(E)$ is the particle speed at energy $E$. Both TSA and VDA SRTs presented in this paper were shifted forwards by the light propagation time from the Sun to Solar Orbiter in order to enable a direct comparison with electromagnetic observations from Solar Orbiter.

We note that for several SEE events, we changed the original flare association based on the results by Papaioannou et al. (in prep.). These events were widespread SEPs, which made the identification of their solar origin ambiguous. This study presents a list of 75 SEP events observed by Solar Orbiter reaching HET energies for both electrons ($\geq$ 1~MeV) and protons ($\geq$~10 MeV), along with a detailed analysis for associating the parent solar sources. The SEE events under consideration are: \textit{2202160445} (C-025-0020), \textit{2203102039} (C-025-0021), \textit{2204200513} (C-025-0028), \textit{2204301745}, and \textit{2208290517}. The number in parentheses corresponds to the identifier used by \cite{2024Dresing}, who compiled a list of 45 SEP events observed by multiple spacecraft in the heliosphere.  In some of these cases, Solar Orbiter was poorly connected to the source, resulting in a delay of several hours between the particle onset and the associated eruption. We note that there may be additional SEE events in our list with incorrectly associated parent solar sources, which would require further detailed analysis. 

\begin{figure}
    \centering
    \includegraphics[width=\linewidth]{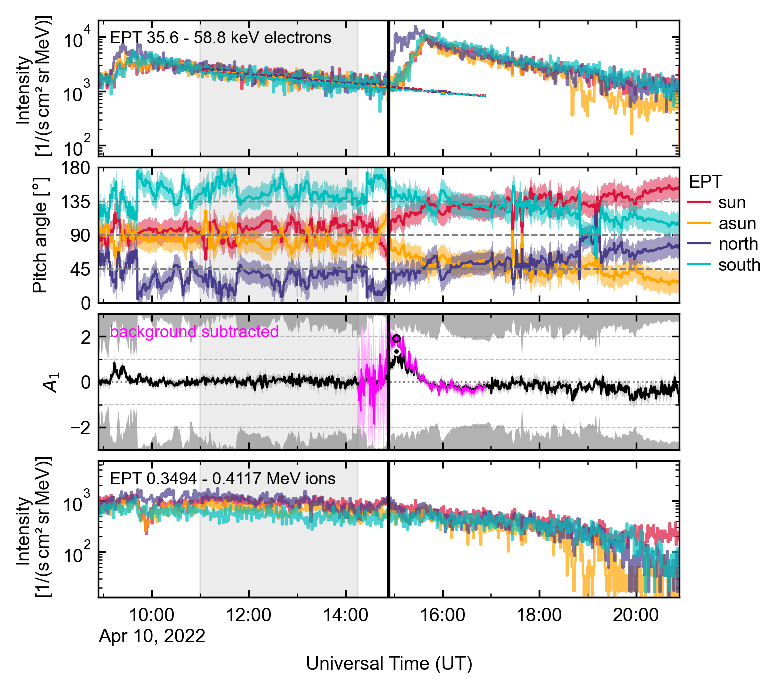}
    \caption{Example plot of EPT observations on 2022 April 10 (event ID \textit{2204101453}). The vertical black line denotes the event onset time. The selected background interval is shaded in grey. Top panel: 35.6--58.8\,keV electron intensities (solid lines) as observed by the four telescopes. The dashed lines show the modelled background intensities during the time interval of interest. Second panel: Pitch-angle ranges covered by the telescopes. Third panel: First-order anisotropy for the measured electron intensities without (black line) and with background subtraction (magenta line) with the 95\% confidence intervals shaded and the peaks denoted as the black and the magenta dots, respectively. The dark grey shading at the top and bottom boundaries denotes anisotropy values that cannot be observed with the given pitch-angle coverage. Bottom panel: 0.349--0.412\,MeV ion intensities as observed by the four telescopes.}
    \label{fig:anisotropy_example}
\end{figure}

\subsubsection{Anisotropies}\label{subsec:anisotropy}
For each SEE event we determined the first-order anisotropy observed by EPT for electrons of 35.6--58.8\,keV, corresponding to the default energy channel used to determine the SEE onset and peak intensities. We used the weighted-sum method by \citet{Bruedern2018} proposed for four-sector measurements:
\begin{equation}
    A_1 = 3\frac{\sum_{i=1}^{N} \delta \mu_i \, \mu_i \, I(\mu_i)}{\sum_{i=1}^{N} \delta \mu_i \, I(\mu_i)} = 3\frac{\sum_{i=1}^{4} \delta \mu_i \, \mu_i \, I(\mu_i)}{\sum_{i=1}^{4} \delta \mu_i \, I(\mu_i)},\label{eq:anisotropy}
\end{equation} where $\mu_i$ is the central pitch-angle cosine of the $i$th telescope, $\delta \mu_i$ is the pitch-angle cosine range of the telescope opening cone, and $I(\mu_i)$ is the observed particle intensity in the $i$th telescope. The classification of the anisotropy was based on the peak first-order anisotropies calculated using background-subtracted intensities. 

As an example, Fig.~\ref{fig:anisotropy_example} shows the SEE event ID \textit{2204101453}. The top panel shows the 35.6-58.8 keV electron intensities (solid lines) observed by the four EPT telescopes. In order to apply background subtraction, we first determine the potentially time-varying background. Therefore, we fit the observations in the background window (highlighted in grey) with exponential functions with a constant decay time and chose the model with the lowest reduced $\chi^2$. As the telescopes gathered observations at different pitch-angles, we also determined whether the background was better modelled with pitch-angle-dependent models. If this was the case, we applied a pitch-angle dependent background subtraction. Finally, we extrapolated the background model (dashed lines in the top panel) forwards in time up to 2 hours after the SEE event onset. The third panel shows the first-order anisotropies calculated using Eq.~(\ref{eq:anisotropy}) both for the measured intensities (black line) and for the background-subtracted intensities (magenta).  We used bootstrapping to estimate their uncertainties considering Poisson errors of the observed counting rates and uncertainties in the background fits. Finally, the ``peak'' anisotropies (plotted as the black and magenta dots) were determined at the moment of time which maximises ${A_1}^2/(A^{\mathrm{97.5th}}_1-A^{\mathrm{2.5th}}_1)$, where $A^{\mathrm{97.5th}}_1$ and $A^{\mathrm{2.5th}}_1$ are the 97.5th and the 2.5th percentiles of ${A_1}$ resulting from the bootstrap analysis, respectively. All events were checked by eye, and instants of peak anisotropy were manually fixed if necessary. Notably, this background removal significantly affects the determined anisotropy for many of the SEE events, and we conclude that without a proper background subtraction SEE event anisotropies are often underestimated.

In the catalogue, we encode the peak anisotropy information as one of the following categories, which are based on the absolute values of the peak anisotropy during the early phase of an event: $0\leq |A_1| < 1$: small; $1\leq |A_1| < 2$: medium; $2\leq |A_1| \leq 3$: large. These categories are used in Fig.~\ref{fig:aniso}, which is discussed in Sect.~\ref{sec:results_ani}.

\subsubsection{Composition}\label{subsec:composition}

\begin{figure*}[t!]
\sidecaption
  \includegraphics[width=12cm]{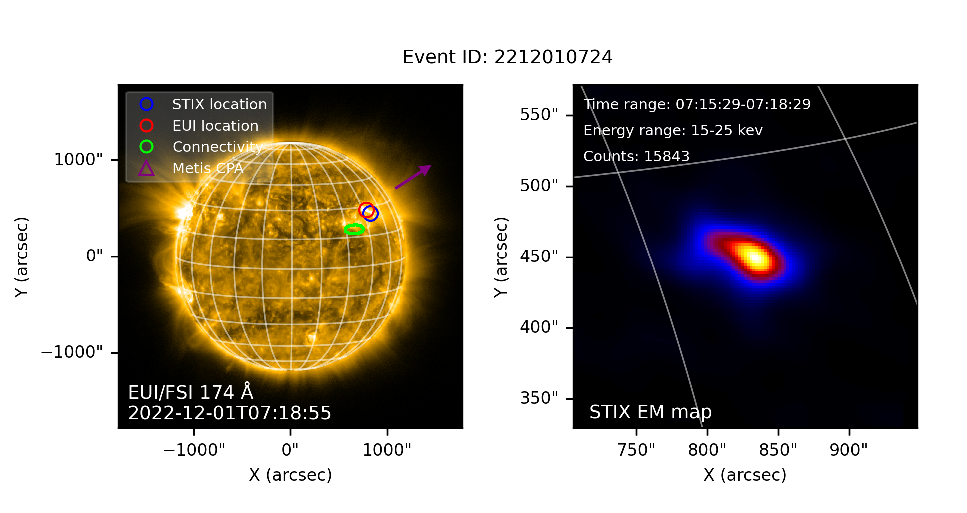}
    \caption{Example plot showing flare locations and magnetic connectivity for an event on 2022 December 1 (event ID \textit{2212010724}). Left panel: Full-disc EUV image in the 174~\AA\ band provided by EUI/FSI. Overplotted are the flare locations derived from STIX and EUI, shown as blue and red circles, respectively. The green ellipse indicates the footpoint of magnetic connectivity provided by the IRAP tool. The size of the ellipse reflects the uncertainties in longitude and latitude. The purple arrow indicates the central position angle of a possibly associated CME observed by Metis. Right panel: X-ray source reconstructed from STIX data with the EM algorithm.}
    \label{fig:stix_eui_example}
\end{figure*}

The EPD/SIS instrument identifies the type of particle intensity increase using the elemental and isotopic composition for the ions H--Fe.  Impulsive SEP events often produce ion intensities only in the range below $\approx$1--2 MeV/nucleon, and typical transit times to the spacecraft are five hours or more at 1 au and proportionally smaller at closer heliocentric distances.  This timescale limits injection timing accuracy to $\pm$15 minutes at best and is subject to systematic errors larger than those of electrons due to scattering and IMF meandering. Additionally, impulsive events often occur in series \citep[e.g.][]{2021Bucik, 2023Kouloumvakos,Lario2024}, as an active region (AR) remains magnetically connected to the spacecraft, and multiple events may not be resolvable from the ion data.  For these reasons, SIS was used here to set a general context for the events observed by the STIX and other EPD instruments.  

In order to allow for the longer transit times for low-energy ions, the particle composition was measured several hours after the electron SRT, with the exact time depending on the spacecraft’s heliocentric distance. The range of delays was ~1.5-5 hours.  The composition in the energy range 0.4--2.0 MeV/nucleon was categorised according to the following characteristics: impulsive: statistically significant $^3$He present and/or Fe/O ratio $\approx$1; gradual: $^3$He not present and Fe/O $\approx$0.1; intermediate: $^3$He injection during a period with otherwise gradual composition, as during the decay of a large SEP event or during a Corotating Interaction Region (CIR); unknown: insufficient statistics to determine the composition.
In case of multiple unresolved electron events, we associated the ion injection with the electron event with the highest peak intensity.
 
We also recorded whether the event was part of a series of $^3$He-rich events, which is defined as the presence of impulsive composition and enhancement of $^3$He and/or high Fe/O that lasted more than 24 hours. Finally, we checked whether the event had a dispersive onset, which means that there was a clear solar injection of heavy ions (mass > 10 amu) showing arrival times inversely proportional to velocity.

\subsection{Associated flares}
\label{sec:flare}

\subsubsection{X-ray flares}
\label{sec:flare_stix}

We used the functionalities provided by the STIX Data Center\footnote{\url{https://datacenter.stix.i4ds.net}} \citep{Xiao2023} to associate STIX flares with the SEE events recorded by EPD. For each event we plotted the STIX quicklook light curves around the inferred electron SRTs. The light curves represent count rates with a temporal cadence of 4~s and are accumulated over the broad energy bands of 4--10, 10--15, 15--25, 25--50, and 50--84~keV (see bottom panel of Fig.~\ref{fig:epd_stix_rpw_example} for an example). We selected the closest STIX flare to the derived SEE SRT, with the additional constraint that the flare must have peaked before the onset of the electron event at Solar Orbiter. In case both TSA and VDA SRTs were available, the latter was used, as it was considered to be more reliable. In four events, the flare association was based on multi-spacecraft observations (see Sect.~\ref{sec:epdtiming}). We adopted our reference time as the time of the main STIX peak at the highest energy range where a flare signature could be clearly seen. In case the flare showed multiple peaks, we also recorded the time of the peak that was closest to the inferred SEE SRT (using VDA if available) and the time of the peak closest to the onset of the associated type III radio burst (cf. Sect.\ref{sec:type3}).

We also defined a confidence level for the association, ranging from 1 to 3. The level is high (1) when only one STIX flare with a single peak in the nonthermal range can be associated with the SEE event (or in the case of multiple peaks, if one of them is clearly favoured); medium (2) when one STIX flare with several peaks corresponds to the EPD event; low (3) when the association is ambiguous (more than one flare potentially associated, SRT delay of more than 1.5 hours, or no STIX data) or no enhancement in X-ray flux is visible at all.

The flare location was determined by reconstructing the X-ray sources using the Expectation Maximization imaging algorithm adapted for STIX \citep[EM;][]{Massa_2019}, which is implemented in the STIX ground software package that is part of the Solar Software IDL (SSWIDL) framework\footnote{\url{https://www.lmsal.com/solarsoft/}}. Using this count-based method, we derived maps of the HXR emission in helioprojective Cartesian coordinates (HPC). We caution that this indirect imaging method cannot provide reliable results when two (or more) ARs flare simultaneously on different parts of the solar disc, which may happen with increasing solar activity.

For each event, we selected the highest possible energy range where enough counts were detected, aiming to image the non-thermal emission that traces the footpoints in the chromosphere rather than thermal emission from the hot plasma in the corona. However, for weak events (e.g. B-class flares), we often used the lower energy range 4--10~keV, so that sufficient counts were available for image reconstruction. This energy range is usually dominated by thermal emission.
The time range was optimised for each event in order to obtain enough counts while covering the main nonthermal HXR peak, which represents the period of the most efficient electron acceleration. An example of a HXR map is shown in the right panel of Fig.~\ref{fig:stix_eui_example}. Finally, a 2D Gaussian function was fitted to the image to measure the location of each source centroid. In the case of multiple footpoints, the source coordinates refer to the one with the highest intensity. This location is overplotted on a full-disc EUI image in the left panel of Fig.~\ref{fig:stix_eui_example}, together with the positions of the EUI event (Sect.~\ref{sec:flare_eui}) and the connectivity footpoint (Sect.~\ref{sec:connect}), as well as the direction of the associated CME observed by Metis (Sect.~\ref{sec:cme}).

STIX was also used to parametrise the X-ray flare importance, since the GOES SXR fluxes that are usually used to define this were only available for about half of the events. Therefore, we used an estimated GOES peak flux that was derived from the measured STIX count rate in the 4-10~keV band \cite[see][]{Xiao2023}.  Note that in the catalogue we only give the actual GOES classes for events that were listed in the solar event reports issued by NOAA's Space Weather Prediction Center. 

\subsubsection{EUV flares and eruptions}
\label{sec:flare_eui}

The identification of SEE-associated flares and eruptive phenomena in EUV was performed using the EUI instrument, including data from FSI in wavelengths 304~\AA~and 174~\AA -- representing chromospheric and coronal origin -- and, when available, from HRI. When FSI data were used, we specified whether the observations were conducted in full-disc mode or in coronagraph mode. 

Flare identification using EUI data involved two steps. Initially, we manually identified solar flares and eruptive events through visual inspection, focussing on the time near the main STIX flare peak. The position of the flare candidate was measured using the JHelioviewer\footnote{\url{https://www.jhelioviewer.org}} software \citep{Muller2017}. We obtained up to three potentially associated EUI flares for each individual SEE event. While all positions were recorded in the catalogue, the EUI source closest to the STIX flare was always adopted as the primary one. 
The left panel of Fig.~\ref{fig:stix_eui_example} shows an example of an EUI-FSI image with the overplotted source locations. In this case, there is only one EUI flare, which is consistent with the STIX source position.

After identifying the EUI events, we associated a NOAA AR number to each event. This was done by first loading EUI FSI 174/304 images, along with continuum and magnetogram data from the Helioseismic and Magnetic Imager \citep[HMI;][]{Scherrer2012} on board the Solar Dynamics Observatory \citep[SDO;][]{Pesnell2012}, into JHelioviewer to accurately identify and track the flare’s source regions. The track feature in JHelioviewer was then used to follow the AR progression until it rotated to the Earth-visible side of the solar disc. For ARs located beyond the east limb, the data were advanced forwards in time to capture the region as it appeared on the visible disc, while for those beyond the west limb, the timeline was reversed by rotating backwards in time. This procedure ensured that the assigned NOAA AR numbers correctly corresponded to the region’s appearance as seen from Earth.

Furthermore, for each pre-selected region, we categorised the eruption type in flares, erupting filaments, loop openings, jets, and fan-like eruptions. The last three can be defined as follows: (1) Loop-like: a large loop structure is seen at the beginning of the eruption, with either one or both legs anchored on the Sun. The eruption takes place when the top of the loop opens, or the legs detaching from the Sun \citep[e.g. ][]{Neupert2001}; (2) Jet-like: narrow, collimated plasma ejections that typically originate from small-scale reconnection events in the solar corona. They are often associated with open or quasi-open magnetic field lines that allow plasma to escape in a directed, beam-like manner. Jets tend to be elongated and maintain a well-defined structure as they propagate. They frequently exhibit an inverted-Y morphology \citep[e.g. ][]{Raouafi2016}; (3) Fan-like: these eruptions involve plasma spreading out over a broader area, often following a fan-shaped pattern. These are usually linked to fan-spine magnetic topologies, where reconnection at a null point results in plasma being expelled in a wider, less collimated way compared to a jet. Instead of a single, narrow column, the plasma expands in multiple directions \citep[e.g. ][]{Cheng2023}.

\subsection{Associated type III radio bursts}
\label{sec:type3}

For each SEE event detected by EPD, we used RPW data to automatically search for type III radio bursts that occurred within a time window of 45 minutes (or longer if the STIX peak time of the associated flare was earlier than these 45 minutes) before and after the EPD SRT (as determined by VDA, if available, or TSA otherwise). The search was first conducted between 3 MHz and 5 MHz, namely around 4 MHz. If at least half of the HFR frequencies in this range showed a flux larger than a certain threshold for at least two time steps (one time step is typically between two and ten seconds), the RPW onset time was recorded as the first time for which this condition was verified. At each frequency, the threshold was set as the minimum value between the median flux plus three times its standard deviation and four times the median flux. The latter threshold usually dominates (i.e. it tends to be the minimum value) in case of larger bursts creating a large standard deviation. This choice was made because the background (noise) spectrum can vary significantly over time due to spacecraft interference. 

The same procedure was then applied around 1 MHz, between 0.8 MHz and 1.2 MHz, where the radio flux is statistically maximal for type III bursts \citep{Sasikumar2022}. If the onset time at 1 MHz followed the onset time at 4 MHz by less than two minutes, only one burst was identified, and the RPW onset time determined between 3 and 5 MHz was adopted. If not, a burst was identified for each detection at each frequency range, unless the onset time occurred after the EPD onset time. Several bursts can be detected at each frequency range, and the final retained RPW onset time was determined by visual inspection to be as close as possible to the SRT as determined from VDA or TSA analysis. This visual inspection was also used to check for the presence of type~II radio bursts.

\begin{figure}
    \centering
     \includegraphics[width=\linewidth]{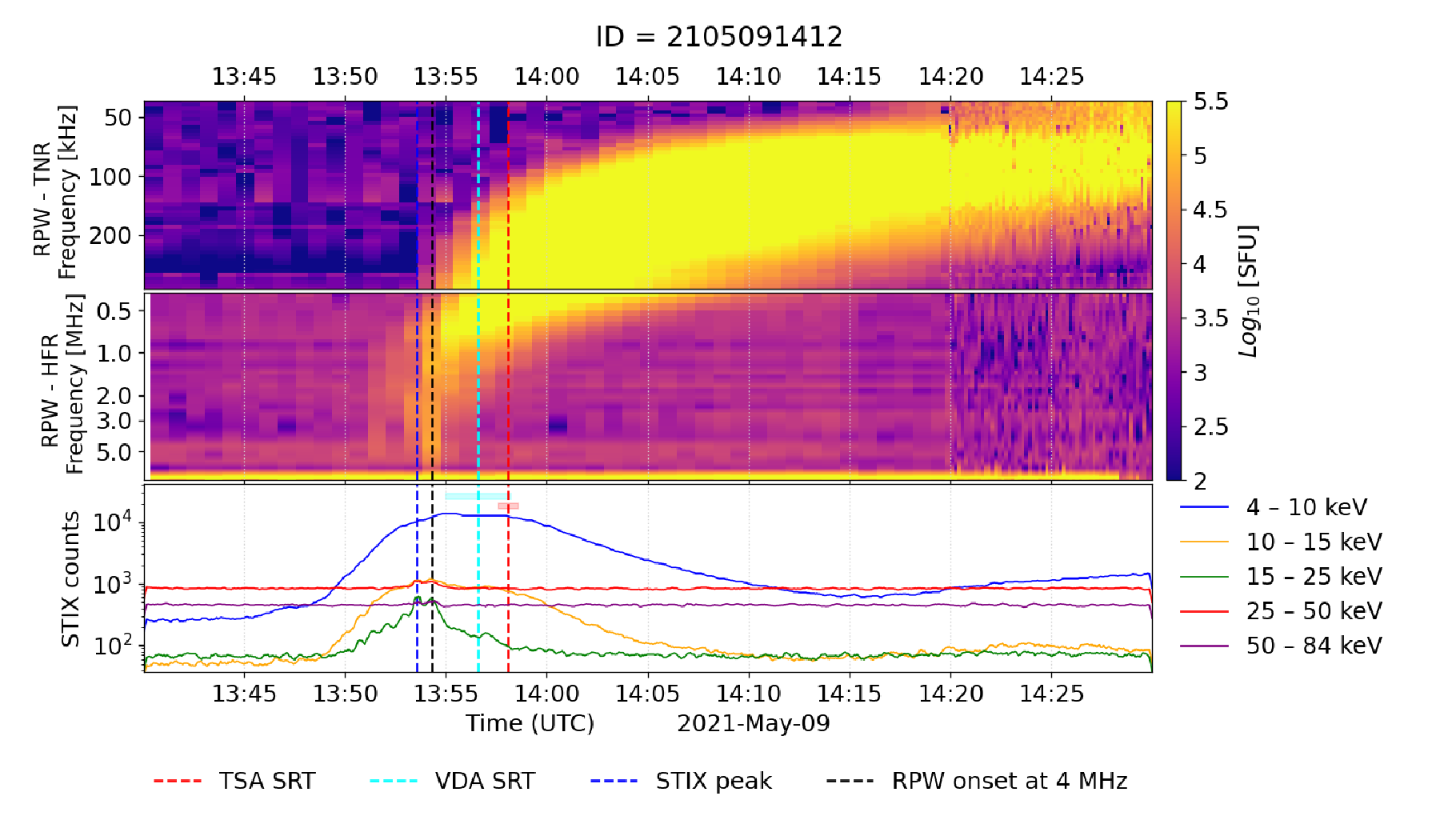}
    \caption{Example of a combined RPW-STIX plot for an event on 2021 May 9 (event ID \textit{2105091412}). The two upper panels show the measured radio flux in solar flux units (SFU) versus time and frequency measured by RPW/TNR and RPW/HFR. The bottom panel shows the X-ray flux in counts per 4 sec as measured by STIX. Vertical dashed lines indicate the following times: electron SRTs derived from EPD data by TSA and VDA analysis, time of maximum X-ray flux from STIX, and the RPW onset times (details given in the main text) detected at 4 MHz or 1 MHz. Horizontal shaded areas at the top of the bottom panel indicate the uncertainties of the TSA and VDA SRTs following the same colour code.}
    \label{fig:rpw_stix_example}
\end{figure}

An example of this procedure is illustrated in Fig.~\ref{fig:rpw_stix_example} which shows RPW (top panel) and STIX data (bottom panel) for the SEE event ID \textit{2105091412}, together with the relevant times indicated by vertical dotted lines (see figure caption for a description). The uncertainty on the VDA SRT is the $\pm 1\sigma$ error on the estimated fit parameter, while the uncertainty on the TSA time corresponds to the integration time of the EPD data. Both uncertainties are indicated in the figure as horizontal shaded areas in the bottom panel. 

\begin{figure*}[]
\centerline{
\includegraphics[width=0.25\textwidth]{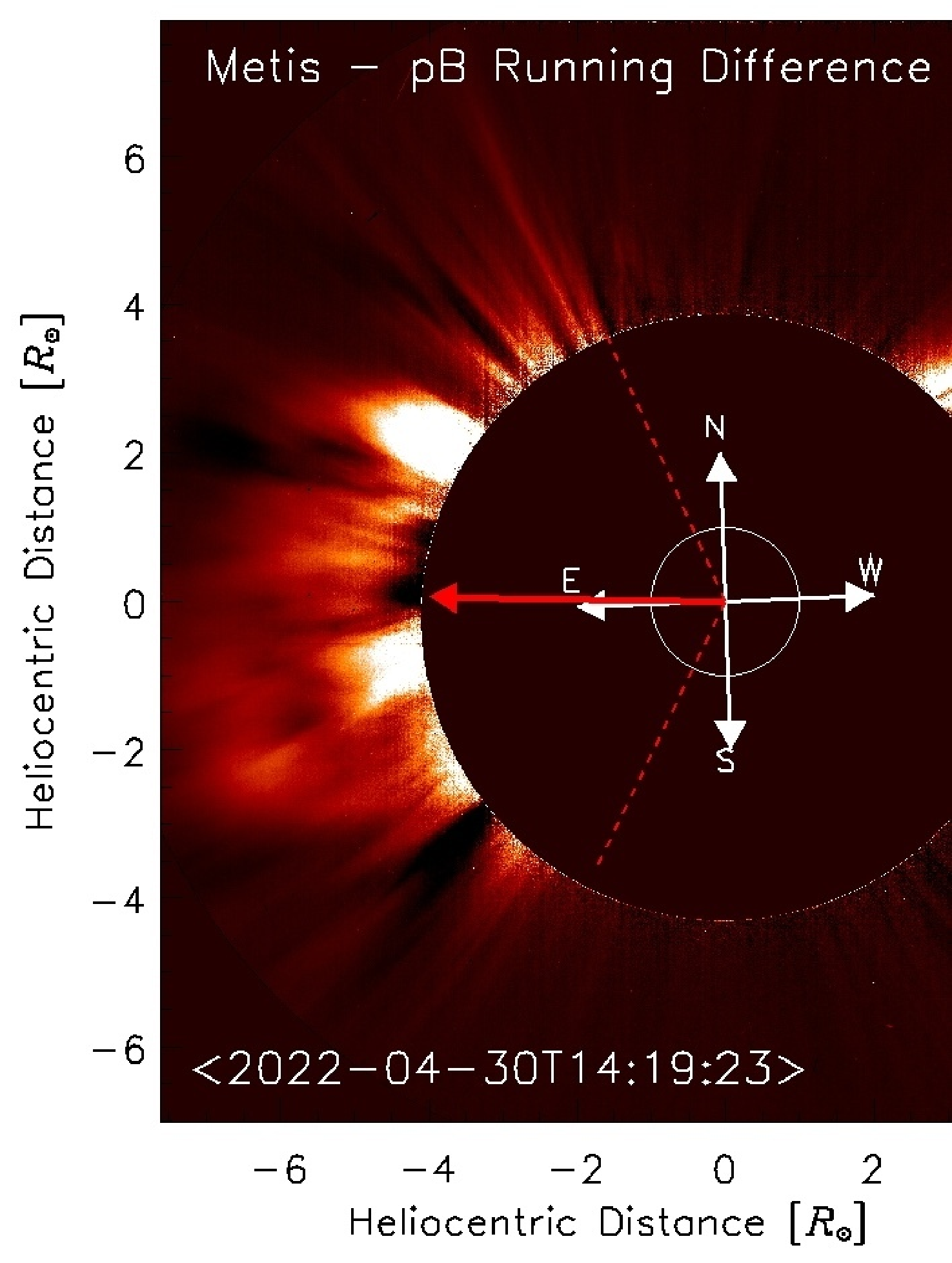}
\hspace*{.1\textwidth}
\includegraphics[width=0.4\textwidth]{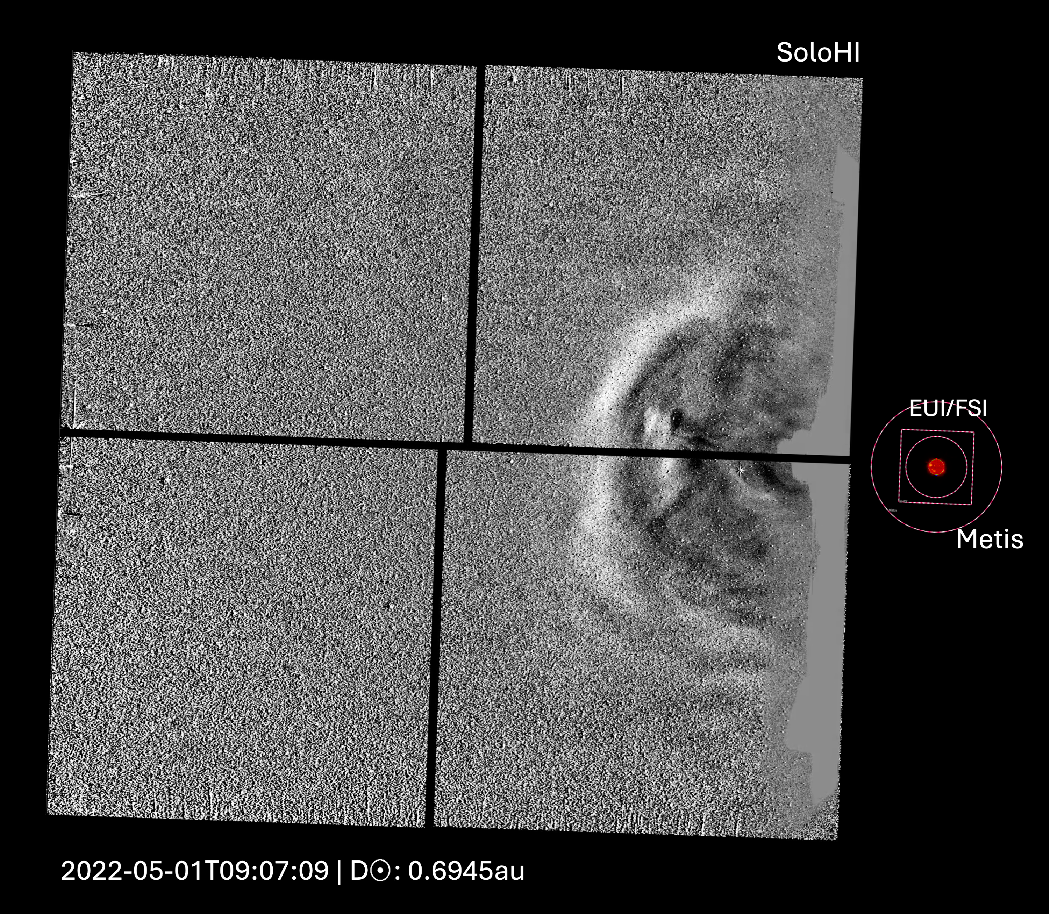}
}
\vspace{-0.29\textwidth}   
\centerline{\Large \bf 
\hspace{0.07 \textwidth} \color{black}{(a)}
\hspace{0.38\textwidth}  \color{white}{(b)}
\hfill}
\vspace{0.28\textwidth}
\color{black}
 \caption{Examples of SEE-associated CMEs (event ID \textit{2204301745}). (a) Running-difference image of the CME observed on 2022 April 30 at 13:19:23 in the VL channel of Metis.
The red arrow represents the central position angle (87$^\circ$) and the dashed red lines  the angular width of the CME.
(b) Running-difference image of the same CME, observed on the following day by SoloHI. The FOVs of Metis and EUI/FSI are delimited with dashed pink lines for comparison.}
     \label{fig:metis_solohi}
\end{figure*}

\subsection{Associated CMEs}
\label{sec:cme}

Solar energetic electron events can be associated with transient phenomena observed in the middle corona, particularly with CMEs in the case of the most intense and energetic SEE events. A CME catalogue\footnote{\url{https://metisarchive.astro.unifi.it/cme/list}} compiled by the Metis team was used to identify CMEs observed in the two Metis channels (i.e. UV and VL) covering distances from 1.7 to 12.9~$R_{\odot}$ depending on Solar Orbiter's heliocentric distance. The Metis catalogue was populated by visually checking Metis image sequences for transients, and recording a range of parameters. The catalogue also provides movies from both channels in terms of total and polarised intensity or running difference images. 

Here, we associated, whenever possible, the SEE events with the CMEs recorded in the Metis CME catalogue. We recorded the CME start and end time, corresponding to the acquisition time of the first and the last image, respectively, in which the transient feature is clearly visible, and the edges of the Metis FOV, which change according to the distance of Solar Orbiter to the Sun. The CMEs were characterised with the central position angle (corresponding to the angle between the CME's central axis and a reference direction, measured counter-clockwise from solar north), the angular width, and the CME speed. The latter was calculated by measuring the position of a selected feature belonging to the CME, at an approximately fixed position angle, in successive images, and applying a linear fit to these plane-of-sky positions. This speed represents a lower limit of the de-projected speed for events propagating out of the plane-of-sky. Fig.~\ref{fig:metis_solohi}a shows an example of a Metis CME with the angular width and central position angle indicated.

To identify the possible association of an event observed by Metis with an SEE event, we used temporal and spatial constraints. Knowing the time of the first CME detection, the position of the inner edge of the Metis FOV, and the estimated propagation speed in the plane of the sky, and assuming constant speed, we could trace back the propagation of the event to determine the launch time of the CME, defined as the time at which the CME leaves the solar surface. This allows us to select a reliable time interval where we verified if a flare associated with an SEE event was identified. For several CMEs (22\%) in the Metis catalogue, it was not possible to determine the speed. One of the reasons for this was the cadence of the Metis observations during the synoptic programme in the period 2020-2023, which was often two hours or more. This made it difficult to follow CMEs in more than one frame and to identify and track the same feature across images to determine their speed. However, this situation is expected to improve from 2023 onward, as the cadence of Metis observations increases. 
Moreover, it is difficult to determine the plane-of-sky speed in both halo and partial halo events, because there are no clear features to track.
It was nevertheless decided to report the existence of these CMEs in this catalogue even if it was not possible to estimate the time when they left the Sun.

In order to correlate the SoloHI observations with the ones provided by the other instruments (e.g. EUI, Metis), we conducted a visual inspection of transients within the SoloHI FOV and selected those that matched the properties previously described (e.g. time, source, direction). Figure~\ref{fig:metis_solohi}b shows an example of a SoloHI CME, highlighting the large FOV as compared to Metis and EUI. As a validation step, we verified the source using the SoloHI catalogue\footnote{\url{https://science.gsfc.nasa.gov/lassos/ICME_catalogs/solohi-catalog.shtml}}, which provides a comprehensive description of each event detected by SoloHI, from its source to 1 au.

\subsection{Interplanetary context}
\label{sec:ip_context}

The properties of the solar wind are governed by various large-scale phenomena that originate from or erupt from the solar corona. The primary large-scale structures found in the solar wind that can be measured in situ include interplanetary CMEs (ICMEs), stream interaction regions (SIRs), IP shocks, and the heliospheric current sheet (HCS).

When intercepted in situ, ICMEs typically consist of three main parts: the sheath, which is a turbulent, compressed plasma region produced by the interaction between the original flux rope of the ICME and the upstream solar wind; the magnetic obstacle or ejecta, which is the core part of the ICME, typically characterised by a low plasma beta, a coherent magnetic field structure, and often a flux rope configuration; and the post-CME which may exhibit properties of both the magnetic obstacle and the ambient solar wind, reflecting the complex dynamics following the passage of the ICME. The HCS results from  regions of oppositely directed magnetic fields at the solar surface.
We also included small-scale flux ropes (SS FRs) identified as  rotations of the magnetic field typically lasting for less than six hours. They might be associated with ICMEs or with in situ reconnections, usually in the vicinity of the HCS.

Stream interaction regions are formed by the interaction between fast solar wind streams, typically emanating from coronal holes, and the slower solar wind ahead of them. This interaction creates a region of compressed plasma and magnetic fields. We can often identify three main parts in the in situ measurements during the transit of a SIR over the spacecraft: the compression region, which is the ambient plasma where the fast solar wind catches up with and compresses the slower solar wind upstream. This leads to an increase in plasma density and magnetic field strength; the stream interface (SI), which corresponds to the boundary separating the fast solar wind from the slower wind, typically close to the highest total pressure value of the interval \citep{Gosling1978}; and the
rarefaction region, which is the trailing part downstream the SI, where the fast solar wind has passed through, resulting in a lower density and magnetic field strength compared to the compression region. The IP shocks are discontinuities in the solar wind speed, density, temperature, and magnetic field. IP shocks can be classified into different types based on their properties (slow, fast, forward, reverse).

The magnetic field topology and characteristics of the solar wind play a crucial role in the propagation of SEEs. For instance, if SEEs are injected directly within a CME, the low-turbulence conditions of the plasma in the magnetic obstacle may result in less scattering of the particles. However, the travel time could increase because the magnetic field lines within the CME might be longer than those in the ambient solar wind \citep{Richardson1996, Gomez-Herrero2017, Wimmer2023,Rodriguez-Garcia2025}. Generally, SEEs propagate through the ambient solar wind where they may encounter various large-scale structures along their path, which can either hinder or facilitate their travel to the observer \citep[e.g.][]{Lario2022}.

In order to identity the presence of large-scale structures that might affect the SEE transport, we used SWA/PAS and MAG data. The selected time interval for identification spans from one day before the arrival of the particles (which may have an influence in the plasma conditions at the arrival of the electrons), as determined by EPD, to three days after their arrival (interval that likely encompasses the plasma regions through which the electrons travelled). The criteria for identifying the different structures are similar to those used in previous studies, such as \cite{sir3,Lan_Jian_1,Lan_Jian_2}, \cite{richardson2010}, and \cite{nieves2018}. An approximate time difference between the SEE onset and the encounter with each structure is provided in the catalogue (Appendix \ref{sec:contents}). 

Although the solar wind is highly variable and propagates at a speed different from that of the SEEs, this approach provides a rough estimate of the conditions encountered by the SEEs from the Sun to Solar Orbiter. In addition, in order to provide more contextual information, in the case that an SEE event occurs in the ambient solar wind, the catalogue tags if the solar wind is fast or slow as based on a threshold of 450 km/s. In these cases, the magnetic polarity (positive for outwards-directed or negative for inwards-directed fields) is also indicated, assuming that the magnetic footpoint is constrained within $\pm60$ degrees of a nominal Parker spiral in the ecliptic plane.

\begin{figure}
    \centering
     \includegraphics[width=\linewidth]{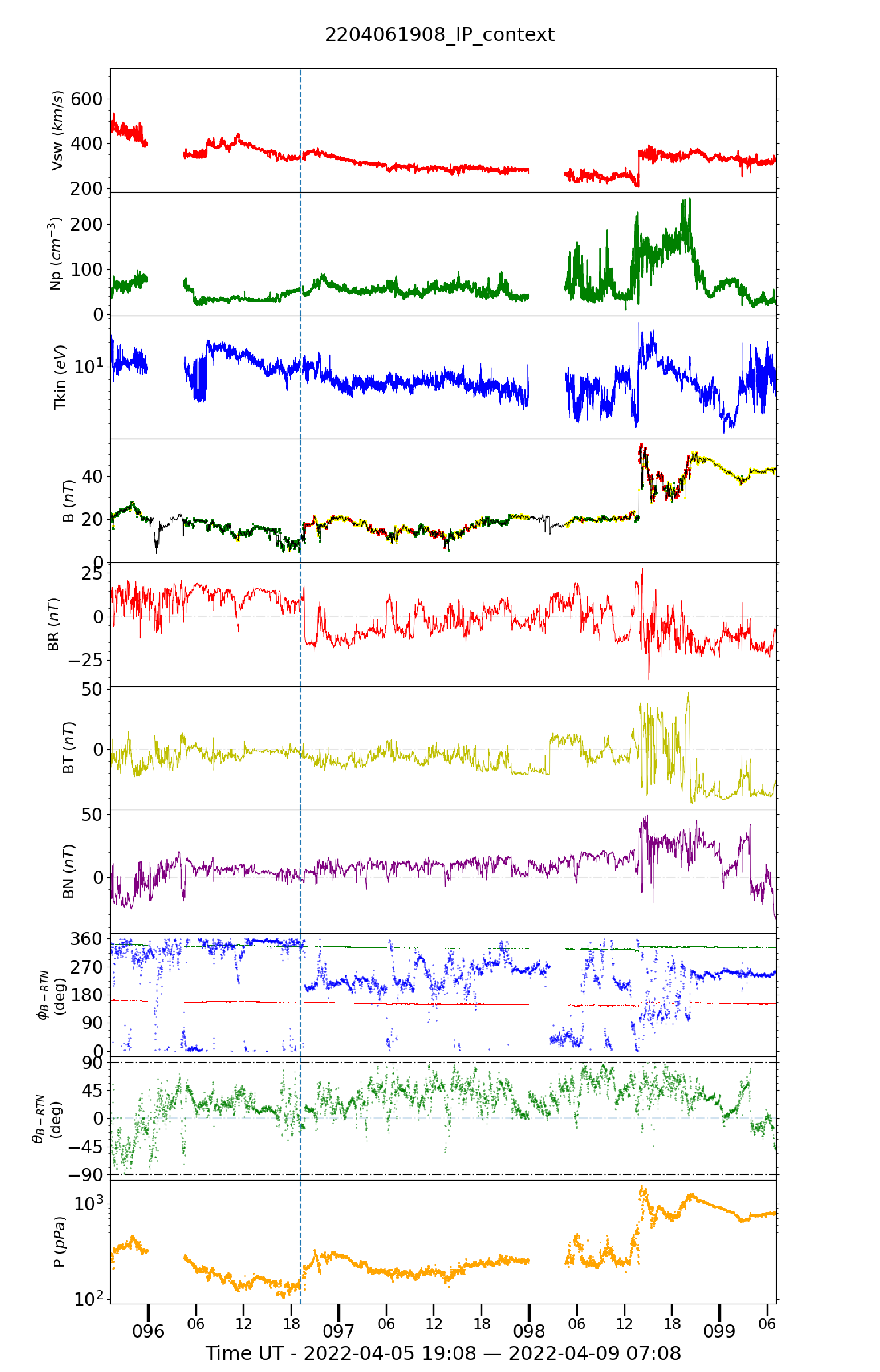}
    \caption{Example plot showing the conditions of the IP medium during a time window ranging from 1 day before until 2.5 days after an SEE onset on 2022 April 6 (shown as the vertical dashed line; event ID \textit{2204061908}). From top to bottom: solar wind proton speed, proton density, proton temperature, IP magnetic field magnitude accompanied by its polarity (red, negative; green, positive; yellow, ambiguous), RTN magnetic field separated components, magnetic field azimuthal angle in the  coordinate system complemented with the two possible nominal Parker spiral angles (red, negative; green, positive. Derived from proton speed), IP magnetic field latitudinal angle, and total pressure.}
    \label{fig:ip_context_example}
\end{figure}

Figure \ref{fig:ip_context_example} shows the Solar Orbiter plasma and magnetic field measurements for the SEE event ID \textit{2204061436}. The represented period spans from 2024 April 5 at 19:08 UTC to 2022 April 9 at 07:08 UTC. The different panels show, from top to bottom, the bulk solar wind speed, proton density, proton kinetic temperature, magnetic field strength (colours represent the polarity: green, positive; red, negative; yellow, undetermined), magnetic radial, tangential, and normal (RTN) components, magnetic field azimuthal angle accompanied by two possible nominal Parker spiral angles (red, negative; green, positive) as derived from the proton speed, IP magnetic field latitudinal angle in the RTN coordinate system, and total pressure. The SEE onset time is marked by the vertical dashed blue line. This onset occurs close to a crossing of the HCS, which can be identified by the sudden change in polarity of the magnetic field (almost 180 degrees). Approximately 42 hours after the onset, a sudden increase in the bulk speed, density, temperature, and magnetic field indicates the cross of an IP shock. The presence and proximity to the onset of these large-scale structures are tagged in the produced catalogue.

Due to interactions between different structures, the IP conditions can become complex and difficult to analyse. In some cases, some SEE events can simultaneously be detected in multiple IP structures or conditions. Such events are tagged as complex events. 

\begin{figure*}[]
\centerline{
\hspace*{0.02\textwidth}
\includegraphics[width=0.58\textwidth]{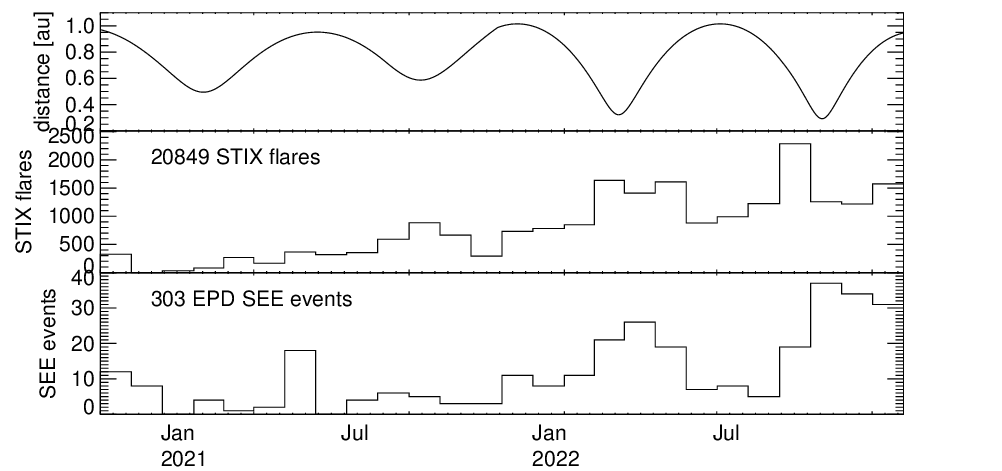}
\hspace*{-0.02\textwidth}
\includegraphics[width=0.4\textwidth]{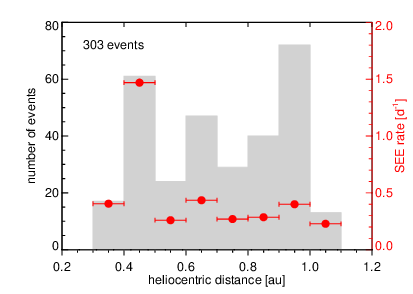}
}
\vspace{-0.27\textwidth}   
\centerline{\Large \bf   
\hspace{0.0 \textwidth}  \color{black}{(a)}
\hspace{0.55\textwidth}  \color{black}{(b)}
\hfill}
\vspace{0.25\textwidth}
\caption{(a) Overview of the time period considered in this study (November 2020 to December 2022). Top: heliocentric distance of Solar Orbiter. Middle: monthly number of X-ray flares recorded by STIX. Bottom: monthly number of SEE events detected by EPD. (b): SEE events according to heliocentric distances of Solar Orbiter at the time of detection. The daily SEE rates for all distance bins are overplotted as red circles. The red error bars show the distance bins over which the rates were calculated.}
     \label{fig:see_rate}
\end{figure*}

\subsection{Magnetic connectivity}
\label{sec:connect}

In order to estimate the magnetic connectivity of Solar Orbiter to the solar surface we used the Magnetic Connectivity Tool \footnote{\url{http://connect-tool.irap.omp.eu/}}. This online tool uses a combination of coronal and heliospheric magnetic field models to estimate the source position of the solar wind and energetic particles measured by different spacecraft \citep{Rouillard2020}. For the extensive list of solar events analysed in the present study we used the tool in its simplest setup where the coronal magnetic field is given by a Potential Field Source Surface (PFSS) reconstruction \citep{Wiegelmann_2012} extending from the solar surface (1 $R_{\odot}$) to a defined source surface (here 2.5 $R_{\odot}$) and beyond this source surface to the Solar Orbiter the IMF is modelled as a Parker spiral. The photospheric magnetic field was provided by Air Force Data Assimilative Photospheric flux Transport (ADAPT) model \citep{Arge_2010}. For each event, the tool automatically selects the best PFSS reconstructions-magnetograms combination according to the \citet{Poirier2021} method.

This represents one of the simplest approaches to derive magnetic connectivity and carries some inherent uncertainty since the IMF geometry may deviate significantly from the nominal Parker spiral due to solar wind turbulence and large-scale IP disturbances. In addition, the PFSS model assumes that the solar corona is current free and in its lowest energy state which is questionable during high levels of solar activity. In order to estimate the uncertainty in the magnetic field tracing in the IP medium, the tool considers a distribution of connectivity points at the source surface around the nominal Parker spiral connection point and then traces hundreds of field lines down to the surface of the Sun \citep{Rouillard2020, Poirier2021}. This simple derivation of uncertainties in the mapping is particularly useful when the Parker spiral connects in the vicinity of sector boundaries and other separatrices that can map to widely separated regions at the solar surface.  

The connectivity tool was used to obtain the magnetic footpoints for the Solar Orbiter spacecraft, using as input parameter the solar wind speed measured in situ by SWA on Solar Orbiter. When this measurement was not available, we considered a set of magnetic footpoints derived from assuming slow (400 km/s) solar wind speed. Each magnetic footpoint was given a probability density according to the hundred magnetic field lines associated with the previously described technique, so we were able to determine the area with the highest probability of location.
In our catalogue, we provide the longitude and latitude of the centre of this area in Carrington coordinates, as well as their uncertainty corresponding to the longitudinal and latitudinal width of the area. In addition, a connectivity confidence level is given. It was calculated using the scatter of the footpoints, their total width and height, and the total probability density of the area, and ranges from 1 (high confidence) to 4 (low confidence).

\section{Results}
\label{sec:results}

\subsection{Event occurrence}

The observing time range considered for this first data release starts on 2020~November~17 (the first day on which both EPD and STIX were operational) and ends on 2022~December~31. During this interval, EPD was operational for 744 days (corresponding to a duty cycle of 96\%) and detected a total of 303 SEE events fulfilling the selection criteria specified in Sect.~\ref{sec:see}.

Figure~\ref{fig:see_rate}a provides an overview of this time period. From the top, the figure shows the heliocentric distance of the spacecraft, the monthly number of X-ray flares recorded by STIX, and the monthly number of selected SEE events detected by EPD. The growing number of STIX flares results from the increasing level of solar activity during the rising phase of solar cycle 25. This is also reflected by the increase in SEE events over time. While the general trend is similar, there is no precise correlation of the monthly SEE rate with the flaring rate. The monthly SEE rate is highly intermittent and ranges from zero to 37 events. This probably reflects the fact that the magnetic connectivity of the solar source to the spacecraft is a crucial prerequisite for an SEE detection, while the flares are detected all over the visible hemisphere.

We compared this result with the statistical study by \cite{Wang2012}, which comprises 1191 SEE events observed at 1 au by the Plasma and Energetic Particle Investigation \citep[3DP;][]{1995Lin3DPWind} instrument on board Wind  \citep{1997Ogilvie}, in the energy range 0.1-300 keV, covering the whole solar cycle 23. They report yearly SEE rates ranging from 12 in solar minimum to 192 in solar maximum, which are significantly lower than the rates provided by EPD. For comparison, EPD detected 226 SEE events in 2022, which was still during the rising phase of cycle 25, when the activity level was clearly below the maximum of cycle 23.

However, we note that the selection criteria used by \cite{Wang2012} differed from ours, as they only included SEE events with velocity dispersion in their list. Considering this, the comparative yearly rate for similar sunspot numbers is approximately 170 SEE events in 2022 (this study) versus around 120 in 1998 (\cite{Wang2012}). However, it is important to note that a significant fraction of the events in 2022 were concentrated in short time periods. The higher rate observed in this study could then be attributed to several factors, including the higher sensitivity of the EPD instrument to discriminate event signatures from the residual background of the preceding events.

Figure~\ref{fig:see_rate}b shows a histogram of the heliocentric distances of Solar Orbiter at the times of the detected SEE events. Generally, more events were detected at larger distances, reflecting the fact that the spacecraft spends more time farther from the Sun due to its elliptical orbit. Nevertheless, 77 events (25\%) were recorded at distances closer than 0.5 au. We note that half of the events in the prominent peak in the histogram between 0.4 and 0.5 au were contributed by a single series of events occurring within only four days in late October 2022. This again demonstrates the highly intermittent nature of the observed SEE activity. 

In Fig.~\ref{fig:see_rate}b, we also show the daily SEE rates for all distance bins (plotted as red circles). Thus, when accounting for the time spent within the various distance ranges, we obtain a more uniform distribution, with the exception of the outlier at 0.4-0.5 au discussed above. Generally, the SEE rate does not vary systematically over distance and remains at a level of 0.3$\pm$0.1 events per day. 

\subsection{Interplanetary context}
\label{sec:res_context}

The IP context in which the SEE events occur can significantly influence their properties, such as onset and rise times, peak intensities, and/or anisotropies. The catalogue compiles the closest temporal approach of different large-scale structures with respect to the electron intensity onset, within a range of $-24$~hr to $+60$~hr, as derived from plasma and magnetic field properties observed by Solar Orbiter (Sect.~\ref{sec:ip_context}).

\begin{figure}[]
\centering
  \resizebox{.73\hsize}{!}{\includegraphics{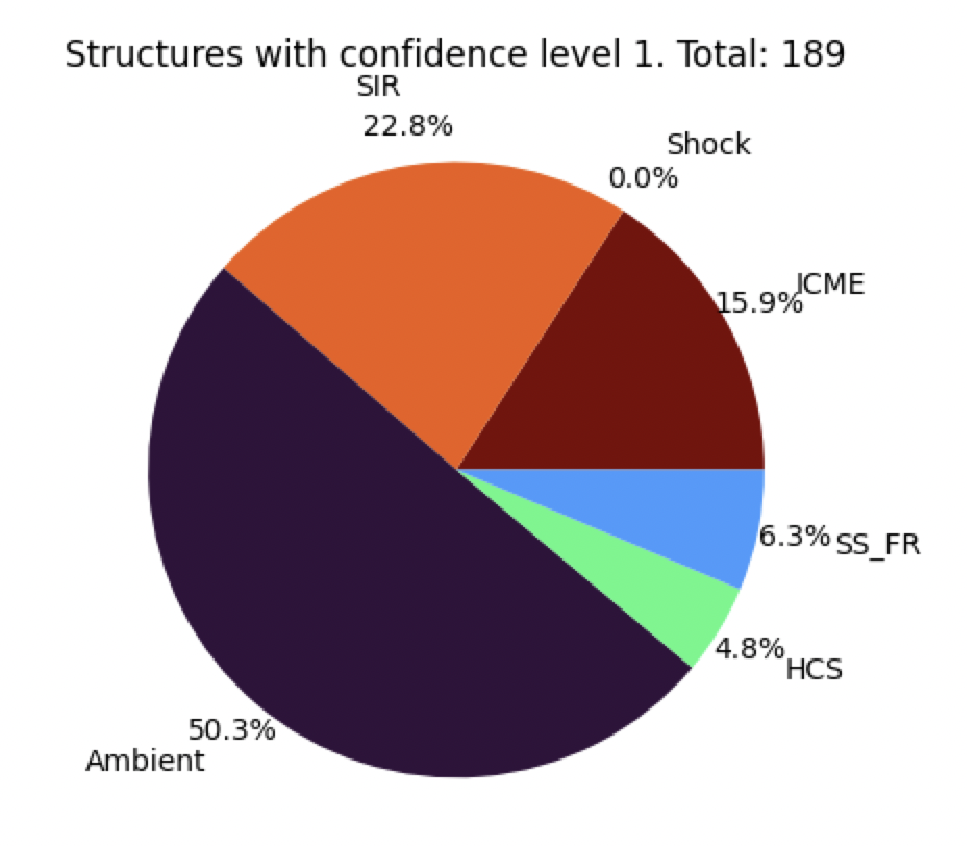}}
     \caption{Relative numbers of SEEs occurring during the transit of different large-scale structures with a confidence level of 1. See text for more details.}
     \label{fig:ip_str}
\end{figure}

As an overview, Fig.~\ref{fig:ip_str} illustrates the distribution of those solar wind conditions where the catalogued SEE events occur (i.e. the identified large-scale structures that were crossing the spacecraft location during the SEE event). These structures are not necessarily associated with the SEE origin, but may have an impact on the particle propagation. The majority of the events (approximately 50\%) take place in ambient solar wind, while the remaining events are associated with various large-scale solar wind structures. 

When the category of the IP structure was extremely unclear, or when measurements were insufficient, the confidence level assigned to the structure was set to at least 2. In cases where the determination was highly uncertain, the confidence level was increased to 3. Fig.~\ref{fig:ip_str} only shows those cases with clear IP context association (confidence level 1).

\subsection{Composition}

\begin{figure}[ttt]
\centering
  \resizebox{1.0\hsize}{!}{\includegraphics{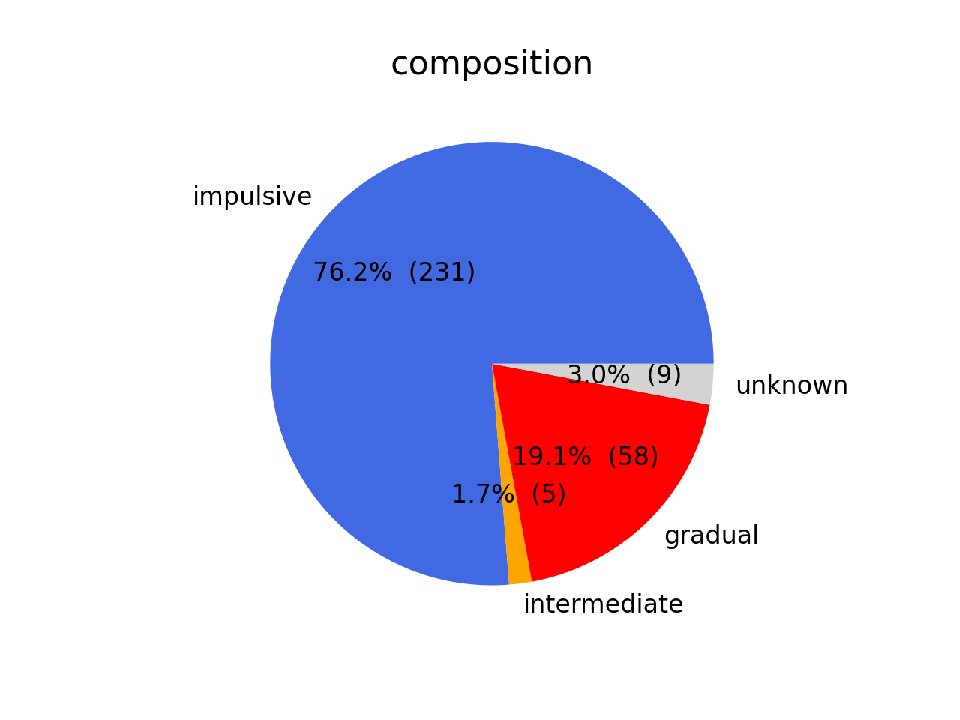}}
  \vspace{-0.06\textwidth}
     \caption{Relative numbers of SEE events according to the composition of the associated energetic ions.}
     \label{fig:compo}
\end{figure}

The pie chart in Fig.~\ref{fig:compo} shows the number of events according to their ion composition measured by EPD-SIS, which was possible in the vast majority of cases (97\%). The composition could not be measured in only nine events, either due to insufficient counts or because SIS was turned off. Events were classified as impulsive if there was statistically significant $^3$He present, and/or Fe/O$\sim$1; gradual if $^3$He was not present, and Fe/O $\sim$0.1; and intermediate for other cases such as a $^3$He-rich injection during the decay phase of a gradual event or during a CIR. Our sample is clearly dominated by SEE events with an impulsive composition (76\% of all cases). Events with gradual composition make up 19\%, and there are only five events of intermediate composition in the sample. The fraction of impulsive events is almost identical to the results of the \cite{Wang2012} survey of 959 SEE events for which the abundance could be measured, of which 75.6\% had $^3$He/$^4$He > 1\%, similar to the criterion for impulsive event classification in this work. 

\subsection{Anisotropy}\label{sec:results_ani}

\begin{figure*}[h!]
\centerline{
\includegraphics[width=1\textwidth,clip=]{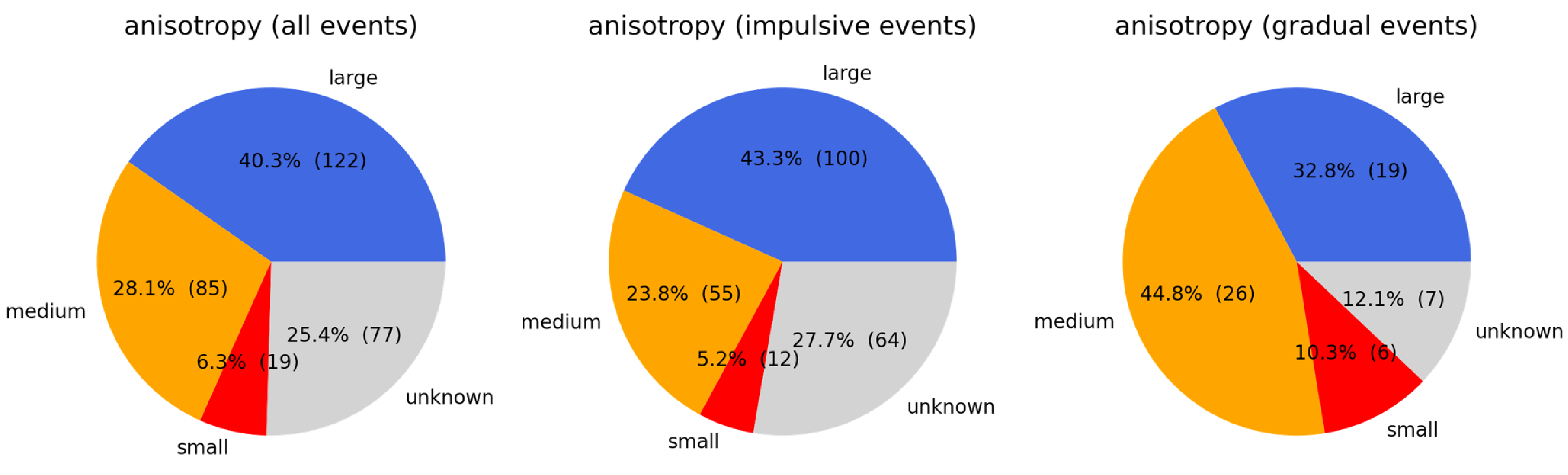}
}
     \caption{Relative numbers of SEE events according to their degree of anisotropy. Left: Anisotropy of all events. Middle: Anisotropy of impulsive events. Right: Anisotropy of gradual events.}
     \label{fig:aniso}
\end{figure*}

The left panel in Fig.~\ref{fig:aniso} shows the number of SEE events according to their degree of anisotropy, which could only be measured in 75\% of events. In the rest, EPT did not detect sufficient counts or magnetic field data were not available. In the majority of events where anisotropy could be measured, at least medium or even larger anisotropies were detected. In only 6\% of all events, the anisotropy was small. We conclude that in the vast majority of SEE events in our sample electrons did not undergo strong scattering processes during their propagation. This strengthens our confidence in the ability to associate solar events with SEEs based on timing.

The middle and right panels of Fig.~\ref{fig:aniso} show the anisotropy fractions for SEE events of impulsive and gradual composition characteristics, respectively. A comparison shows that gradual events tend to have lower levels of anisotropy. Specifically, large anisotropies occur less frequently than in the sample of impulsive events, while small and medium anisotropies are over-represented. When investigating the distributions of rise times as a function of anisotropy (not shown here), we find that events of large and medium anisotropy are both strongly peaked with low rise-time, with medians around 10 min, while events of small anisotropy have a flat distribution with a median of 125 min. This is consistent with poorly magnetically connected events where gradual particle injections, enhanced scattering, and potentially perpendicular diffusion processes may occur \citep[e.g.][]{Zhang2009,Droege2010}. 

\subsection{SEE rise times and intensities}

\begin{figure}[bp]
\centering
  \includegraphics[width=1.\linewidth]{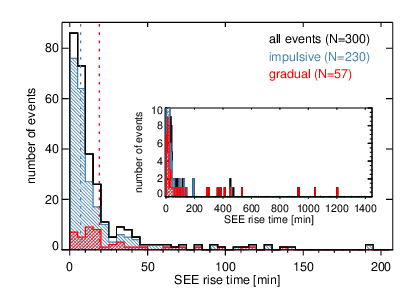}
 \caption{SEE rise times. Impulsive event distributions are shown in blue, gradual ones in red, and the black histogram represents all events. Dotted lines show the medians of the distributions for impulsive and gradual events. While the main panel shows rise times up to 200 min, the inset shows the full range of up to one day.}
     \label{fig:histo_epd_rise}
\end{figure}

Figure~\ref{fig:histo_epd_rise} shows histograms of the SEE rise times, defined as the peak intensity times minus the onset times. The black outline gives the distribution for all events, while the shaded blue and red histograms indicate the distributions of impulsive and gradual events, respectively. The distribution of impulsive events is more strongly peaked at short rise times (i.e. below 20 minutes) than that of gradual events. The median rise times are 7 min and 19 min for impulsive and gradual events, respectively. Gradual events also show a significantly higher number of outliers at very long rise times, which can be seen in the inset in Fig.~\ref{fig:histo_epd_rise}. The maximum rise time was 20 hours. Consequently, the difference in mean rise times is much more significant, namely 18 min versus 143 min for impulsive and gradual events, respectively. This shows that the composition-based classification into impulsive and gradual events is also reflected in their time profiles.

\begin{figure}[t!]
\centerline{\includegraphics[width=1\linewidth]{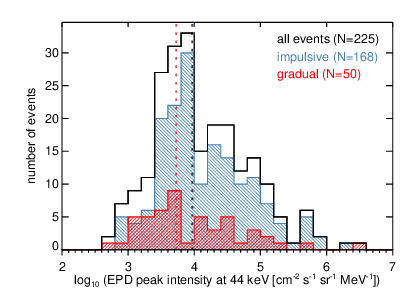}}
\vspace*{-0.32 \textwidth}  
\centerline{\Large \bf 
\hspace{-0.22 \textwidth}
\color{black}{(a)}
\hspace{0.22 \textwidth}
}
\vspace*{0.29 \textwidth} 
\centerline{\includegraphics[width=1\linewidth]{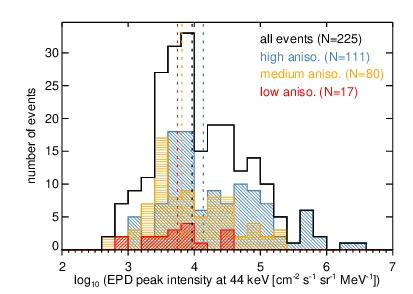}}
\vspace*{-0.32 \textwidth}  
\centerline{\Large \bf 
\hspace{-0.22 \textwidth}
\color{black}{(b)}
\hspace{0.22 \textwidth}
}
\vspace*{0.29 \textwidth}
\centerline{\includegraphics[width=1\linewidth]{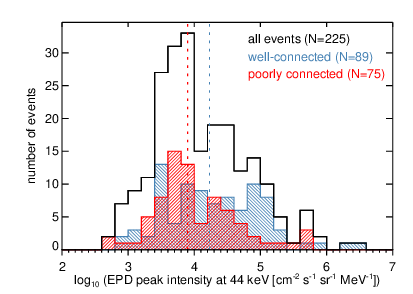}}
\vspace*{-0.32 \textwidth}  
\centerline{\Large \bf 
\hspace{-0.22 \textwidth}
\color{black}{(c)}
\hspace{0.22 \textwidth}
}
\vspace*{0.31 \textwidth} 
 \caption{Results on the SEE peak intensities measured at 44~keV. (a): SEE peak intensities, with impulsive event distributions shown in blue, gradual ones in red, and the black outline representing all events. Dotted lines show the medians of the distributions. (b): Peak intensities for SEEs with large (blue), medium (yellow), and small anisotropy (red). (c): Peak intensities for well-connected SEEs (blue) as opposed to poorly connected ones (red), where well-connected events are defined as having a separation in longitude between the STIX source and the footpoint of magnetic connectivity of less than 20°.}
     \label{fig:flux}
\end{figure}

SEE rise times also depend on anisotropy (not shown here). The median rise times of the electron intensity at 44~keV for events with large, medium, and small anisotropy were 9, 11, and 125 min, respectively. This clearly shows that low-anisotropy events are characterised by significantly longer rise times, which could indicate that diffusion and/or continued particle acceleration plays an important role in these events.

The SEE peak intensities could be measured in 300 events. We adopted the energy band of 35.6--54.2~keV (henceforth referred to its mean of 44~keV) as the default energy at which peak intensities were measured. In 72 events, the electron intensities could only be determined at lower energies, and in three events peak intensities had to be obtained at higher energies. Figure~\ref{fig:flux}a shows the peak intensity distribution for the 225 events where it could be measured at 44~keV. Peak intensities range over four orders of magnitude, from \mbox{$6 \times 10^2$} to \mbox{$4 \times 10^6$}\,cm$^{-2}$\,s$^{-1}$\,sr$^{-1}$\,MeV$^{-1}$. The separate distributions for impulsive and gradual events are also shown, as well as their medians. Note that these two distributions do not differ significantly. Generally, gradual events are assumed to be characterised by higher intensities, which is not the case in the present sample. We believe this is primarily due to the absence of significantly large gradual events recorded between 2020 and 2022, a period when solar activity remained at moderate levels. A living catalogue focussed on large gradual events observed by Solar Orbiter is being prepared by Papaioannou et al., 2025 (in prep). 

The histograms in Fig.~\ref{fig:flux}b show that the peak intensities have a moderate dependence on anisotropy. Highly anisotropic SEEs tend to have higher intensities than events with medium or low anisotropy. Conversely, there are no low-anisotropy events with peak intensities larger than $4 \times 10^4$\,cm$^{-2}$\,s$^{-1}$\,sr$^{-1}$\,MeV$^{-1}$.  When Solar Orbiter observes events with higher peak intensity, the event is more likely impulsive, and hence has a higher anisotropy on average.

We also find a dependency on magnetic connectivity. In Fig.~\ref{fig:flux}c we compare the distribution of peak intensities for well-connected events with the distribution for more widely separated events.  We defined well-connected events as those events that have a longitude separation between the STIX flare location and the footpoint of the predicted connecting magnetic field line of less than 20$^{\circ}$ (see Sect.~\ref{sec:conn}). The well-connected events tend to have larger peak intensities.

With Solar Orbiter, we can test the radial dependency of SEE peak intensities. However, the current sample doesn't show any clear dependence. The likely reason for this is that event-to event variations in peak intensity can cover several orders of magnitude, dominating over radial variations. In order to determine the radial dependence of peak intensities, the same SEE event should be measured at different radial distances by magnetically aligned spacecraft \citep[e.g.][]{Lario2006,Lario2013, 2023Rodriguez-Garcia, 2025Cao}.

\subsection{Association with flares}

\begin{figure}[]
\centering
  \resizebox{1.0\hsize}{!}{\includegraphics{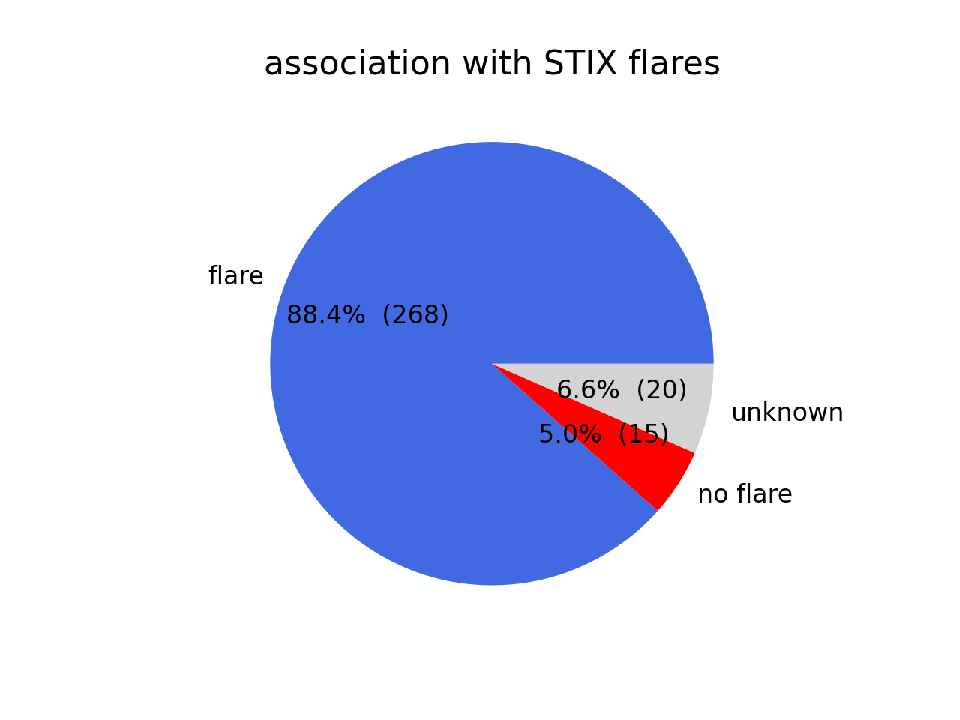}}
    \vspace{-0.06\textwidth}
     \caption{Relative numbers of SEE events according to their association with STIX flares.}
     \label{fig:assoc_stix}
\end{figure}

 Figure~\ref{fig:assoc_stix} shows the association of the SEE events with STIX X-ray flares. STIX data were available for 283 SEE events, with 268 of them being linked to a STIX flare. In 15 events, no enhancement was detected in the STIX light curves. We conclude that the SEE events in our sample are highly associated (>88\%) with X-ray flares. Note that some of the apparent ‘flareless’ events could be originating behind the limb as seen from Solar Orbiter.

\begin{figure}
\centering
 \includegraphics[width=0.48\textwidth]{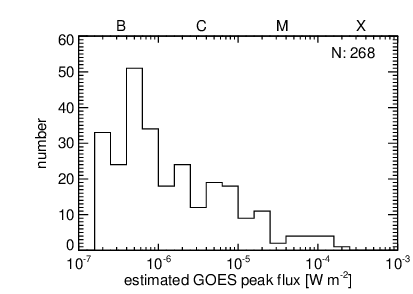} 
  \includegraphics[width=0.48\textwidth]{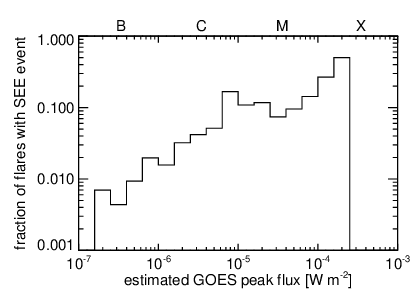} 
\caption{Top: Estimated GOES peak flux of the STIX flares associated with the SEE events. Bottom: Fraction of SEE-associated STIX flares as a function of estimated GOES peak flux.} 
\label{fig:histo_goes}
\end{figure}

\begin{figure*}
\sidecaption
  \includegraphics[width=6cm]{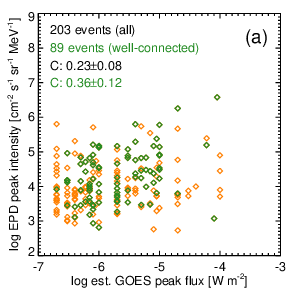}
  \includegraphics[width=6cm]{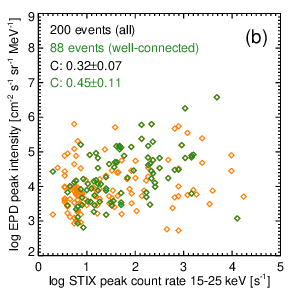}
     \caption{EPD peak intensity at 44~keV plotted against (a) the estimated GOES peak flux and (b) the STIX peak count rate in the 15-25~keV band. The green diamonds correspond to the well-connected events (i.e. events with a separation in longitude between the STIX source and the footpoint of magnetic connectivity of less than 20°), while the rest of the sample is shown with orange diamonds. The plots also indicate the number of events and the correlation coefficients for all events in black, and  for the well-connected events in green.}
     \label{fig:stix_goes_epd_fluxes}
\end{figure*}

The top panel of Fig.~\ref{fig:histo_goes} shows a histogram of the estimated GOES peak flux of the SEE-associated STIX flares. The distribution is clearly dominated by weak flares (142 B-class, 91 C-class, 30 M-class, and 5 X-class flares). To assess the association with SEE events as a function of the flare importance, we normalised this distribution with the corresponding distribution of all STIX flares that were observed during the time period considered in this study, and for which a GOES estimate could be made (20\,632 flares in total). The resulting distribution is shown in the bottom panel of Fig.~\ref{fig:histo_goes}. It is evident that the fraction of SEE-associated flares increases with the GOES peak flux: the fraction is 0.8\% for B-class flares, 3.4\% for C-class, 11\% for M-class, and 29\% for X-class flares. 

We investigated the relationship between flare importance and peak SEE intensity. Figure~\ref{fig:stix_goes_epd_fluxes}a shows a scatter plot of the logarithm of the EPD peak intensity at 44~keV and the logarithm of the estimated GOES peak flux. Taking into account all events, we find a weak correlation of $C=0.23 \pm 0.08$. The correlation is higher ($C=0.36 \pm 0.12$) for well-connected events (shown in green). While this is broadly in agreement with the results of \cite{2023bRodriguez-Garcia} obtained from MESSENGER data, our correlations are somewhat lower. One possible reason could be that Solar Orbiter observes the SEEs at different distances, while SEE peak intensity may decrease as the electrons spread out through the heliosphere. However, scaling the EPD peak intensity by distance to the power of two or three only marginally improved the correlations. 

We proceeded to compare the SEE intensities with a more direct measure of flare importance, namely the STIX count rates in the broad energy bands used in the quicklook light curves (the GOES estimate is based on the STIX count rate at 4-10 keV rescaled to 1~au). Peak count rates as well as background rates for all automatically detected STIX flares were obtained from the STIX Data Center. As expected, the correlation for the background-subtracted 4-10~keV peak count rate is very similar to the GOES estimate, while it improves for 10-15~keV, and reaches its maximum at 15-25~keV with $C=0.32 \pm 0.07$ for all events and $C=0.45 \pm 0.11$ for well-connected ones (see Fig.~\ref{fig:stix_goes_epd_fluxes}b). This behaviour is most probably connected to the transition from thermal to nonthermal X-ray emission. While the 4-10~keV is always dominated by thermal emission of the hot flare plasma, the 10-15~keV range tends to show some contribution from nonthermal emission, while 15-25~keV is usually dominated by nonthermal emission, except in large flares. The slight improvement in the correlation with SEE intensities indicates that the nonthermal electrons in flares are more closely related to the in situ electrons than the thermal flare response that is usually employed to characterise flare strength would suggest, which is consistent with the correlations between the number of accelerated electrons in flares and in situ \citep[][]{Krucker2007,Dresing2021}.

\begin{figure}
\centering
  \resizebox{1.0\hsize}{!}{\includegraphics{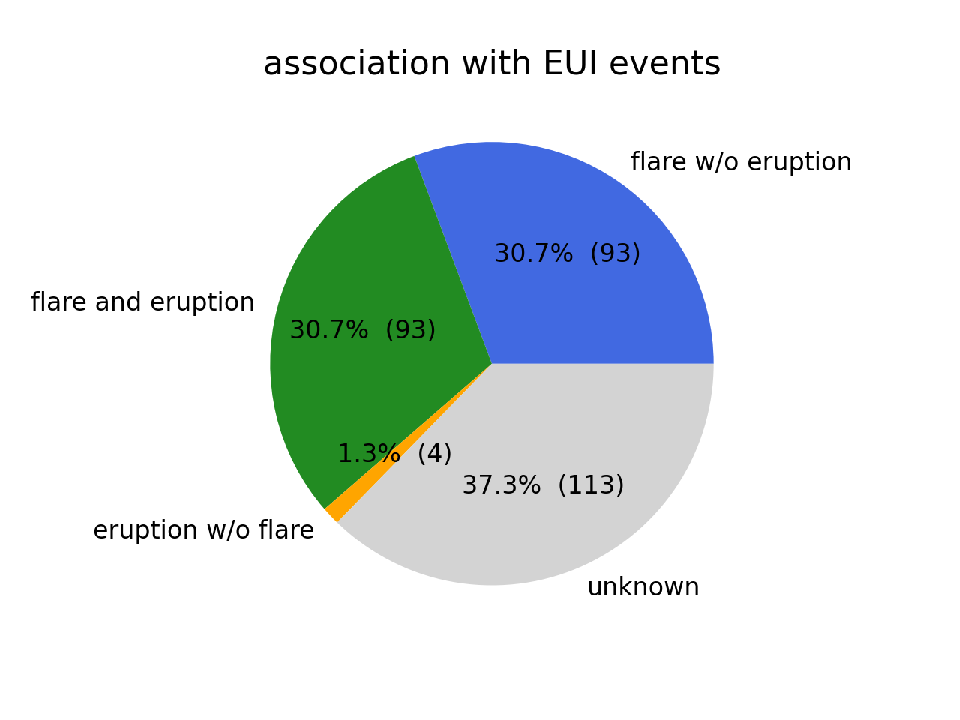}}
    \vspace{-0.06\textwidth}
     \caption{Relative numbers of SEE events according to their association with EUI flares and/or eruptions.}
     \label{fig:assoc_eui}
\end{figure}

\begin{table}[b]
\caption[]{Number of SEE events associated with EUI events of various types (flares and eruptive phenomena).}
\label{tab:eui_events}
\centering
\begin{tabular}{lcc}
            \hline
            \noalign{\smallskip}
            event type      &  number & fraction \\
            \noalign{\smallskip}
            \hline
            \noalign{\smallskip}
            all EUI-associated events &  194 & 100\% \\
            flares &  190 & 98\% \\
            jets &  63 & 32\% \\
            erupting filaments &  14 & 7\% \\
            erupting fans &  13 & 7\% \\
            erupting loops & 8 & 4\% \\
            loop openings & 14 & 7\% \\
            \noalign{\smallskip}
            \hline
\end{tabular}
\end{table}

EUI disc observations were available for 194 events, while EUI was in coronagraphic mode in an additional 30 events. In all cases with disc observations, an EUI flare or eruption could be detected. In 53 of these events, potentially associated EUI signatures were detected at two locations, and in 26 cases at three distinct positions. In the cases with multiple EUI sources, we adopted the one which was closest to the STIX flare as the primary one. This approach was validated by a comparison of the EUI source locations with the footpoint of magnetic connectivity (see Sect.~\ref{sec:conn}). Figure~\ref{fig:assoc_eui} shows the association of the SEE events with flares and eruptive phenomena at the primary EUI location. 

In 31\% of all SEE events, both a flare and indications of various eruptive phenomena were present, the same fraction of events showed flares without eruptions, and eruptions without flares were present in just 1\% of events. However, for more than a third of all events no EUI data were available, either because the instrument was switched off or operated in coronagraphic mode. Table~\ref{tab:eui_events} shows the numbers of EUI flares and different types of eruptions as well as their fraction with respect to the total number of SEE events with detected EUI signature. We note that more than one type of eruption can be associated with a single SEE event. The eruption types are dominated by small-scale features. Narrow jets are the most commonly observed type of eruption related to SEEs (detected in 32\% of the cases with EUI coverage). Erupting fans, which are wider than jets, are observed in 7\% of cases.  Erupting filaments are present in another 7\% of events. We also observe slower eruptions, related to erupting loops (4\%) and loop openings (7\%).

We conclude that most SEE events are associated with X-ray and EUV flares. The fraction of SEE-associated flares clearly increases with peak X-ray flux. Half of the flares where EUI observations were available show eruptive behaviour. We stress that the fraction of eruptive events reported here probably represents a lower limit, since in most events only data from EUI-FSI operating in synoptic mode were available. The limited temporal cadence and spatial resolution do not favour the detection of small-scale impulsive eruptive phenomena such as EUV jets. We plan to do a follow-up study on the events that were visible from Earth using SDO/AIA data to better constrain associated eruptive phenomena. Note that for a series of SEE events in November 2022 that are also covered by CoSEE-Cat, a recent study by \citet{Lario2024} indeed found a high association with EUV jets observed with SDO/AIA.

\subsection{Association with radio bursts}

\begin{figure}
\centering
  \resizebox{1.0\hsize}{!}{\includegraphics{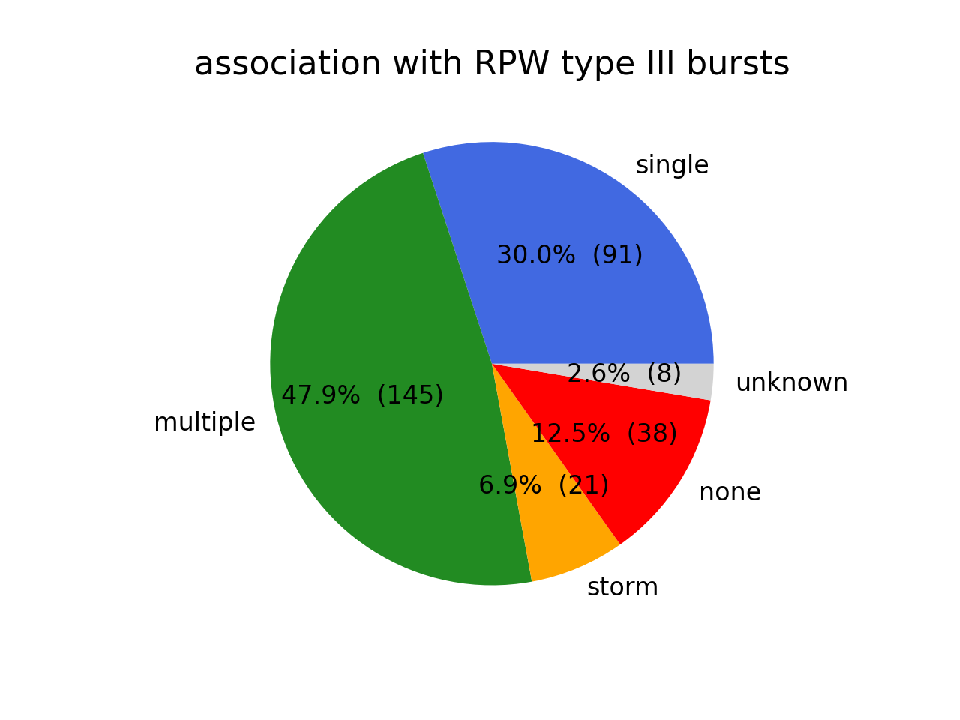}}
    \vspace{-0.06\textwidth}
     \caption{Relative numbers of SEE events according to their association with RPW type III radio bursts.}
     \label{fig:assoc_rpw}
\end{figure}

As is shown in Fig.~\ref{fig:assoc_rpw}, the SEEs events are highly associated with type III radio bursts. In 30\% of the SEEs, a single type III burst was detected, while multiple bursts were present in the largest fraction of events (48\%). 7\% of the events occurred during type III storms. In these cases, no individual bursts could be associated with the SEE event. Finally, 13\% of the SEE events did not show any association with a type III burst.

The low number of non-associated events has to be regarded as an upper limit. When checking dynamic radiospectra from Wind/WAVES \citep{Bougeret1995} , STEREO-A/SWAVES \citep{Bougeret2008} and PSP/FIELDS \citep{Bale2016} for some selected events, we found that in several cases type III bursts were detected by one or several of these spacecraft. Thus SEEs are probably even more highly associated with type III bursts. This will be investigated more systematically in future work.

Additionally, 18 events were associated with IP type II radio bursts detected by RPW, which indicate the presence of shock waves. As with type III bursts, this represents a lower estimate, and additional data sources will have to be checked to derive more comprehensive statistics.

\subsection{Association with CMEs}

\begin{figure}
\centering
  \resizebox{1.0\hsize}{!}{\includegraphics{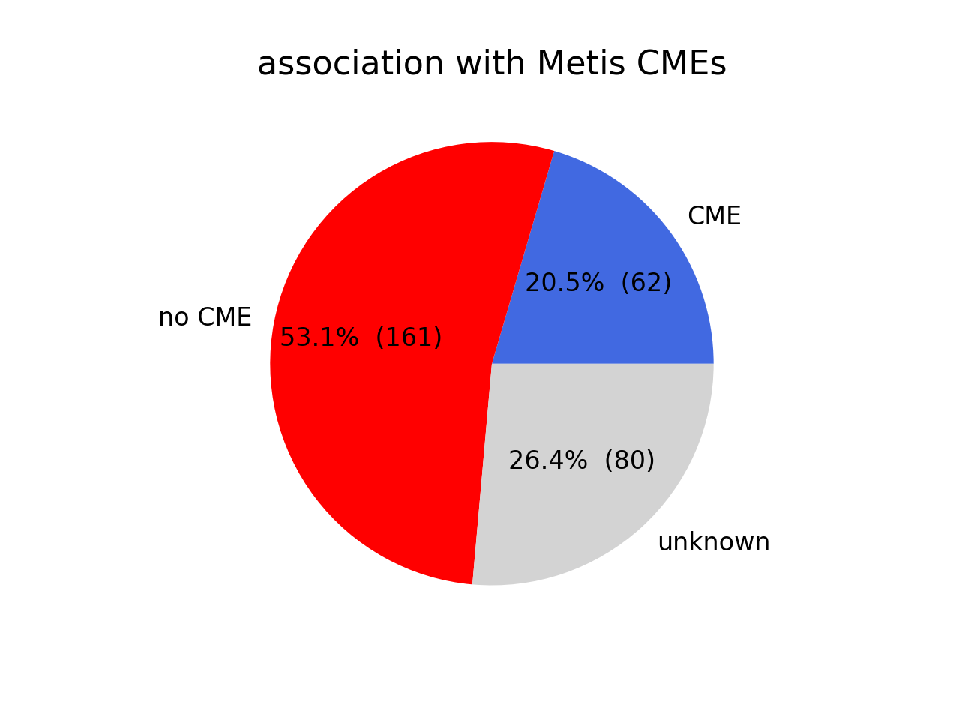}}
 \vspace{-0.1\textwidth}
 \resizebox{1.0\hsize}{!}{\includegraphics{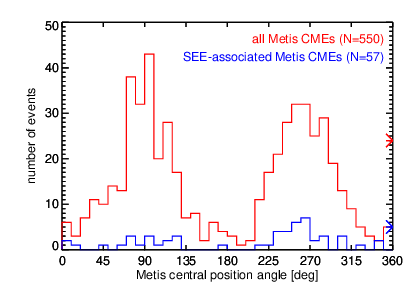}}
  \vspace{0.1\textwidth}
     \caption{Top: Relative numbers of SEE events according to their association with Metis CMEs. Bottom: Distribution of the CME central position angles, for the total sample of observed Metis CMEs (red) and the CMEs associated with SEE events (blue). The asterisks at the right border of the plot show the number of halo CMEs. }
     \label{fig:assoc_metis}
\end{figure}

During the entire period considered in this study, Metis observed a total of 550 CMEs, of which 4\% and 8\% are classified as halo and partial halo, respectively. Across all CMEs, the average latitudinal width is $70\pm77^\circ$ and the average speed is $285\pm148$~km\,s$^{-1}$.
The association between Metis CME observations and SEE events could not be verified for 27\% of the SEE events (80 cases) because Metis was not observing at the time. The upper panel in Fig.\ref{fig:assoc_metis} shows the association of SEE events with CMEs observed by Metis. CMEs were detected in 62 SEE events, and in five of these cases two potentially associated CMEs were observed. The number of CMEs detected by Metis corresponds to 21\% of all SEE events, which increases to 28\% when considering only the periods in which Metis was observing. Conversely, SEE events without an associated CME signature represent 53\% of all cases and 72\% when restricting the analysis to periods of Metis observations. 
Thus, we can infer that slightly less than one third of the SEE events show an association with CMEs detected by Metis. Another notable finding is that the association of gradual events with CMEs is more significant than for impulsive events (29\% vs 19\% for gradual and impulsive events, respectively).

It is possible that very faint or narrow CMEs are present in the observations and could be associated with some SEE events but were not included in the Metis catalogue. This is because the Metis catalogue was created independently of the SEE event catalogue. A more detailed study could involve searching the Metis data for all possible CME signatures associated with each SEE event, but this is beyond the scope of this work.

Note that the number of SEE-associated CMEs is just 57, while 62 SEE events were associated with CMEs. This difference results from the fact that a few CMEs were associated with more than one SEE event. The distribution of central position angles (i.e. the main angle along which the feature propagates) is shown for the 57 SEE-associated CMEs and for the entire set of Metis CMEs in Fig.~\ref{fig:assoc_metis}. Interestingly, CMEs associated with SEE events are almost entirely absent at high latitudes, far from the equatorial belt. Additionally, there is a higher percentage of halo (10\%) and partial halo (29\%) events, and the average angular width of these CMEs ($107 \pm98^\circ$) is greater than that calculated for the entire population of CMEs observed by Metis.
So, the CMEs associated with SEE events are preferably those with a large latitudinal extent and therefore those with ‘halo’ characteristics, which are likely directed towards Solar Orbiter. In terms of plane-of-sky speed, the SEE-associated events do not show a distribution different from that of all Metis events.

\begin{figure*}[h!]
\centerline{
\hspace*{0.07\textwidth}
\includegraphics[width=0.36\textwidth]{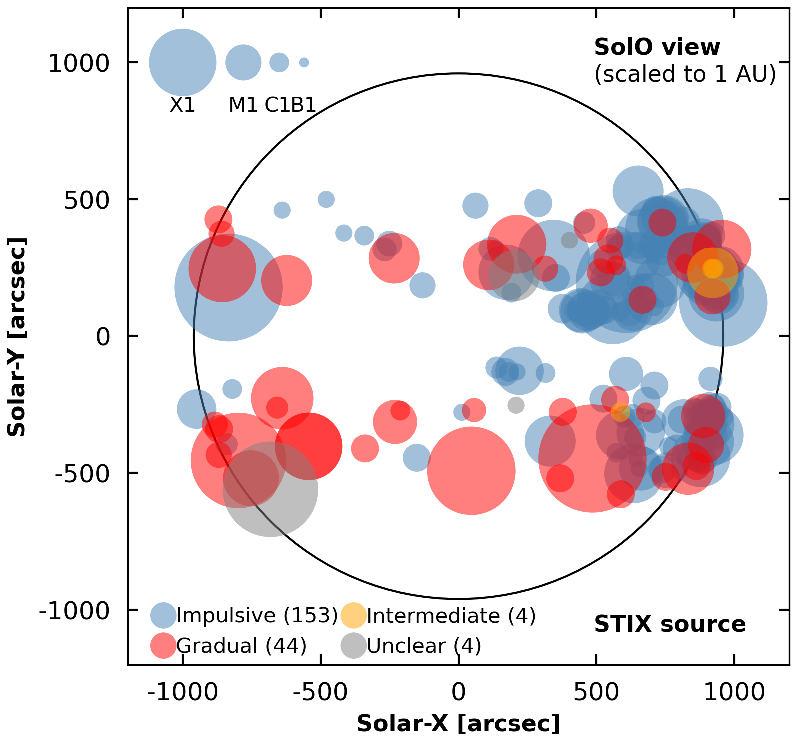}
\hspace*{0.07\textwidth}
\includegraphics[width=0.5\textwidth]{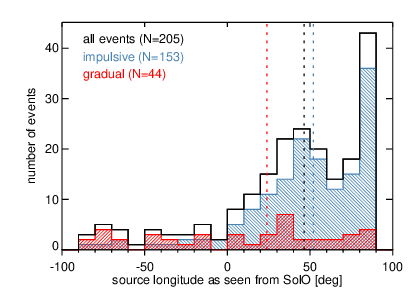}
}
\vspace{-0.35\textwidth}   
\centerline{\Large \bf 
\hspace{0.0 \textwidth}  \color{black}{(a)}
\hspace{0.46\textwidth}  \color{black}{(b)}
\hfill}
\vspace{0.33\textwidth}
\centerline{
\includegraphics[width=0.5\textwidth]{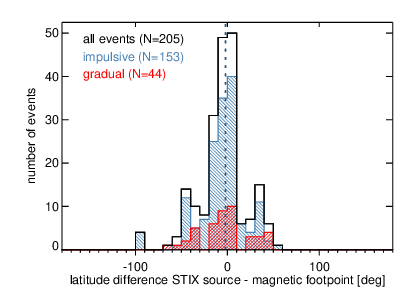}
\hspace*{0.0\textwidth}
\includegraphics[width=0.5\textwidth]{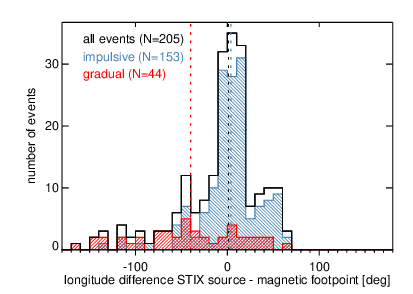}
}
\vspace{-0.35\textwidth}   
\centerline{\Large \bf 
\hspace{0.0 \textwidth}  \color{black}{(c)}
\hspace{0.46\textwidth}  \color{black}{(d)}
\hfill}
\vspace{0.33\textwidth}
 \caption{(a): STIX source positions as seen from Solar Orbiter. All positions have been rescaled to 1 au. The size of the circles corresponds to the GOES peak estimate, while the colours indicate the composition. (b): STIX X-ray source longitude as seen from Solar Orbiter. (c): Heliographic latitude difference between the STIX source and the footpoint of magnetic connectivity. Impulsive event distributions are shown in blue, gradual ones in red, and the black histogram represents all events. Dotted lines show the medians of the distributions for impulsive and gradual events. (d): As in (c), but showing the heliographic longitude difference between the STIX source and the footpoint of magnetic connectivity.}
     \label{fig:stix_pos}
\end{figure*}

During the analysed period, SoloHI detected a total of 111 CMEs. However, only eight of these CMEs could be associated with the SEE events in this catalogue. This outcome was anticipated, as the SoloHI FOV points towards the east of the Sun, thereby missing the majority of events directed towards the west or directly towards Solar Orbiter. However, out of the 62 CMEs detected by Metis and correlated with EPD events, 18 CMEs propagated eastwards and were not detected by SoloHI. Upon careful examination, three primary reasons emerge for this non-detection. Firstly, many of these CMEs occur when Solar Orbiter is beyond 0.75$\,$au, where the cadence and resolution of SoloHI decrease significantly, making it challenging to detect events unless they are sufficiently large. In combination, some of these events are faint and narrow in the Metis FOV, becoming even more difficult to detect when they expand and enter the SoloHI FOV. Additionally, some CMEs propagate at high latitudes, further complicating their identification.
More details about all events detected by SoloHI can be found in the \href{https://science.gsfc.nasa.gov/lassos/ICME_catalogs/solohi-catalog.shtml}{SoloHI catalogue}.

We emphasise that all CME associations are based on timing and do not necessarily imply a causal relation with the SEE events. Establishing such a relation will require a more detailed analysis, such as deriving the CME kinematics in more detail, which will be addressed in a future work. However, it is important to note that even when a CME is not related to a particular SEE event, it may nevertheless have an impact on electron propagation.

\subsection{Source location and magnetic connectivity}
\label{sec:conn}

We considered the locations of the SEE-associated flares and how they relate to the predicted footpoints of the magnetic field lines connecting to Solar Orbiter. For 205 STIX flares (77\% of all STIX flares) we were able to perform imaging and determine a source position. Figure~\ref{fig:stix_pos}a shows the STIX flare positions in helioprojective Cartesian coordinates. All positions have been rescaled to a 1 au viewpoint. The size of the circles reflects the estimated peak GOES flux, while their colour indicates the composition of the associated SEE event. It is evident that the flares originated in the two activity belts and that there is a clear east-west asymmetry in the number of events. In general, we expect Solar Orbiter to be magnetically connected to the western hemisphere as a result of the shape of the Parker spiral, and the impulsive events indeed reflect this. This is also clearly seen in Fig.~\ref{fig:stix_pos}b, where we compare the distributions of the source longitude (in heliographic coordinates) as seen from Solar Orbiter for impulsive and gradual events (shown in blue and red, respectively). While impulsive events are strongly clustered in the western hemisphere with a median longitude of W52, gradual events show a flat distribution with a median of W24. This is very consistent with the results of \cite{Reames1999}. Note that the prominent peak at W80-W90 is partly caused by off-limb events which are recorded as having a longitude of W90. We also have to point out that there are 12 impulsive events with source locations on the eastern hemisphere (five of them beyond E45), which is not assumed to be magnetically well connected to the spacecraft. Some of these events are apparently really poorly connected to Solar Orbiter, which is shown by long SRT delays, long SEE rise times, and lack of velocity dispersion at their onsets. In other cases, there are secondary or tertiary EUI sources that are more consistent with the assumed magnetic connectivity, which implies that the STIX flares were misassociated with the SEE events in these cases.

We also compared the STIX source locations with the predicted connectivity footpoints (Sect.~\ref{sec:connect}). Fig.~\ref{fig:stix_pos}c shows the distribution of the heliographic latitude difference between the STIX source and the connectivity footpoint, while Fig.~\ref{fig:stix_pos}d shows the corresponding longitude differences. We show the distributions for all events and for impulsive and gradual ones. For impulsive events, the longitude difference distribution is strongly peaked around zero, indicating that in most impulsive events, the STIX source location is close to the predicted connectivity footpoint. The same holds for the latitude differences. In contrast, gradual events show a very flat longitude difference distribution with a preference towards negative differences. 

The differences in latitude and longitude are less than 20$^\circ$ in 69\% and 64\% of all impulsive events, respectively. Two nearly symmetrical secondary peaks are observed on both sides of the main latitude peak, as seen in Fig.~\ref{fig:stix_pos}c . They are likely due to the presence of magnetic field loops linking the northern hemisphere to the southern hemisphere of the Sun, which prevent the PFSS model from accurately estimating the magnetic connectivity. The distribution of potential connectivity footpoints is split into two regions, one in each hemisphere, typically between $\pm$[20$^\circ$, 60$^\circ$]. The area with the higher probability predominates, while the location of the STIX source may be found in the opposite hemisphere or somewhere in between. The four points between -90$^\circ$ and -100$^\circ$ represent extreme cases of misidentification: they correspond to a group of events originating from a source identified by STIX around -30$^\circ$ in latitude, while the highest probability for the magnetic footpoints were found between 60$^\circ$ and 70$^\circ$.
With regard to longitude differences, the secondary peak around -50$^\circ$ in Fig.~\ref{fig:stix_pos}d could be the result of an underestimation of the solar wind speed. Indeed, for several cases of the catalogue, the solar wind speed could not be measured, and the generic speed of 400 km/s was used for the PFSS modelling. This may be an underestimate due to the complex dynamic of the solar wind and in particular its acceleration \citep{Dakeyo_2024}. The secondary peak around 60$^\circ$ in Fig.~\ref{fig:stix_pos}d can be partly attributed to SEE events occurring behind the limb and with a longitude reported as 90$^\circ$ (Fig.~\ref{fig:stix_pos}b). The general asymmetry to the negative longitudinal differences for the gradual SEE events simply results from a shift in the distribution of these events over the entire surface of the Sun with respect to the mean longitude connectivity of Solar Orbiter, located around 40--50$^\circ$.

Applying a cut-off of $\pm 20$° in angular distance to filter out apparently misidentified connectivity footpoints, taking the standard deviation of the angular differences, and converting them to the full width at half maximum (FWHM) of a Gaussian distribution, we obtained FWHMs of 26° and 29° for the latitude and longitude differences distributions of impulsive events, respectively. This is consistent with the common assumption that electrons are injected into a cone of 30° average angular extent \citep[e.g.][]{Lin1974, Reames1999,Ho2024}.

We also studied latitude and longitude differences with respect to the EUI flare positions. The distributions for the primary EUI source were qualitatively similar to the results for STIX, but the secondary maxima were significantly higher. This was even more pronounced when we considered the secondary and tertiary EUI sources. This demonstrates that in most cases the primary EUI flare was indeed identified correctly as the source associated with the SEE event.

We conclude from the strongly peaked latitude and longitude difference distributions that impulsive SEE events are launched from localised regions, namely flares or small-scale eruptive events that are magnetically connected to IP space, while the distributions for gradual events are more consistent with extended acceleration or injection regions. Different scenarios exist for the latter case, such as acceleration at a CME-driven shock or magnetic reconnection between a large-scale erupting flux rope and the ambient open field lines \cite[e.g.][]{Klein2005,Klein2024}. 

\begin{figure*}[t!]
\centerline{
\includegraphics[width=0.5\textwidth]{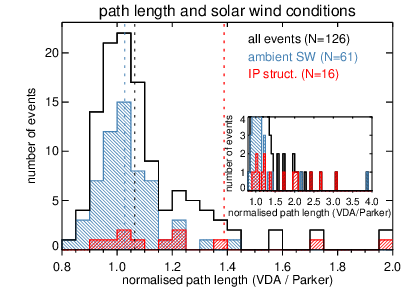}
\hspace*{0.0\textwidth}
\includegraphics[width=0.5\textwidth]{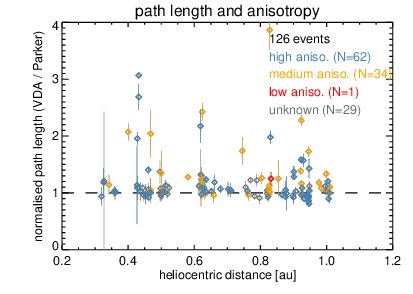}
}
\vspace{-0.35\textwidth}   
\centerline{\Large \bf 
\hspace{0.0 \textwidth}  \color{black}{(a)}
\hspace{0.45\textwidth}  \color{black}{(b)}
\hfill}
\vspace{0.33\textwidth}
 \caption{Results on inferred path lengths. All path lengths derived from VDA fits were normalised by the nominal Parker spiral length. (a) 
 Distribution of normalised path lengths. Events associated with ambient solar wind and IP structures are shown in blue and red, respectively, and the black outline represents all events. Dotted lines show the medians of the distributions. While the main panel shows normalised path length up to 2, the inset shows the full range up to 4. (b) Normalised path lengths plotted versus heliocentric distance. Events with high, medium, and low anisotropy are indicated in blue, yellow, and red, respectively. Events of unknown anisotropy are plotted in grey.}
     \label{fig:path}
\end{figure*}

\subsection{Effective path lengths}
\label{sec:paths}

The VDA fits provided effective IP path lengths in the range from 0.3 to 3.4 au, with a mean relative uncertainty at the 10\% level. In order to account for the changing heliocentric distance and solar wind speed, we focus our analysis on the path lengths that are normalised with the nominal Parker spiral length. The distribution of this parameter is plotted in Fig.~\ref{fig:path}a. It is strongly peaked between 0.9 and 1.2, with a mean of 1.3 and a median of 1.1. Thus, most path lengths do not deviate too strongly from the nominal Parker spiral length, i.e. 73\% of path lengths are within $\pm 20$\% of the nominal length.

We find no significant differences in the path length distributions of impulsive and gradual events, as well as no dependence on anisotropy or the connectivity of the source. However, a significant factor appears to be the IP conditions, specifically, whether an event is observed in the ambient solar wind or within IP structures. Events observed in the ambient solar wind (plotted in blue in Fig.~\ref{fig:path}a) show much better agreement with the nominal Parker spiral than events that are associated with IP structures (plotted in red), with medians of 1.03 and 1.39, respectively. Note that the event numbers in these two samples are reduced since we only considered cases where the confidence level of the IP measurements was rated high.

We note that there is a significant population of outlier events that show large normalised path lengths; for example, eight events with path lengths longer than twice the nominal Parker spiral. Figure~\ref{fig:path}b shows the normalised path lengths as a function of heliocentric distance, colour-coded for anisotropy. The plot shows that the outliers do not depend on distance or anisotropy: they can occur close to the Sun as well as far away, and can be both of medium or large anisotropy. We also find no influence of composition, source connectivity, or IP complexity. Therefore, more detailed case studies are required to explain these extraordinarily long paths. Such studies have already been performed by \citet{Wimmer2023} for events \textit{2204091141} and \textit{2204091152}, and by \citet{Rodriguez-Garcia2025} for event \textit{2201200639}. Both studies concluded that the long paths were due to propagation of the particles within a magnetic flux rope in the context of an ICME. Of the five other events with long normalised path lengths, the associations were as follows: one with complex IMF conditions and a post-CME structure, two events with complex IMF conditions only, one event with an SIR rarefaction region and finally, one event with the HCS.

\begin{figure*}[t!]
\centerline{
\includegraphics[width=0.5\textwidth]{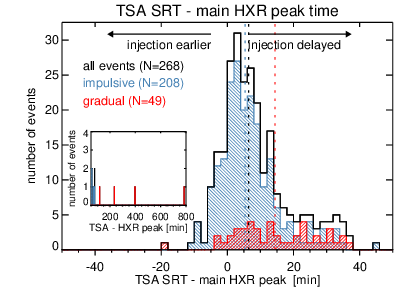}
\hspace*{0.0\textwidth}
\includegraphics[width=0.5\textwidth]{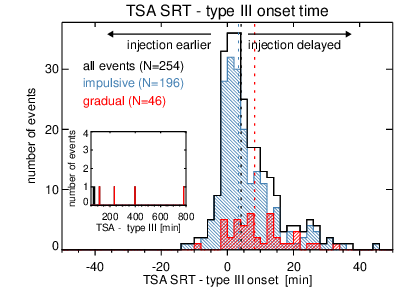}
}
\vspace{-0.35\textwidth}   
\centerline{\Large \bf 
\hspace{0.0 \textwidth}  \color{black}{(a)}
\hspace{0.46\textwidth}  \color{black}{(b)}
\hfill}
\vspace{0.33\textwidth}
\centerline{
\includegraphics[width=0.5\textwidth]{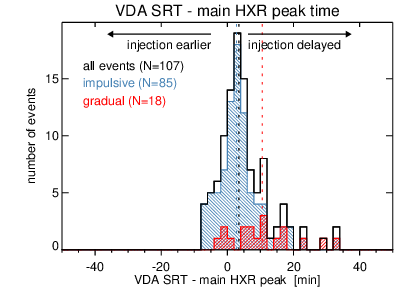}
\hspace*{0.0\textwidth}
\includegraphics[width=0.5\textwidth]{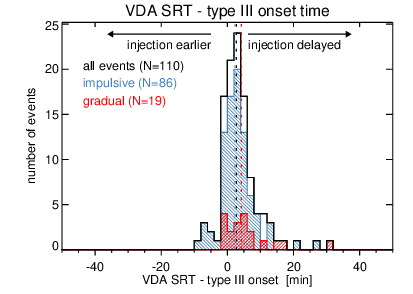}
}
\vspace{-0.35\textwidth}   
\centerline{\Large \bf 
\hspace{0.0 \textwidth}  \color{black}{(c)}
\hspace{0.46\textwidth}  \color{black}{(d)}
\hfill}
\vspace{0.33\textwidth}
 \caption{SRTs of SEEs relative to the times of the main nonthermal STIX peak and the type~III burst onset. (a): TSA SRTs relative to main nonthermal STIX peak. (b): TSA SRTs relative to type III onset. (c): VDA SRTs relative to main nonthermal STIX peak. (d): VDA SRTs relative to type III onset. Impulsive events are shown in blue, gradual ones in red, and the black histogram outlines represent all events. Dotted lines show the medians of the distributions for impulsive and gradual events.}
     \label{fig:histo_stix_tdiff}
\end{figure*}

\begin{figure*}[t]
\centering
\includegraphics[width=.48\linewidth]{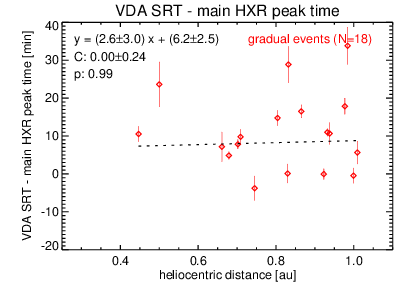}
\includegraphics[width=.48\linewidth]{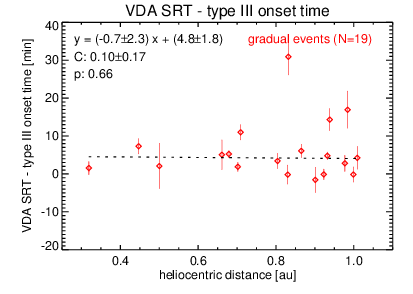}
\includegraphics[width=.48\linewidth]{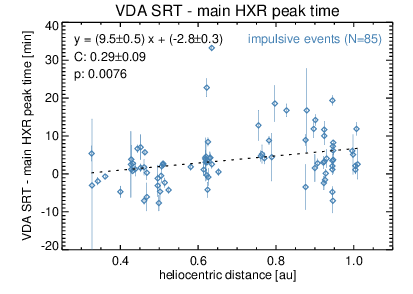}
\includegraphics[width=.48\linewidth]{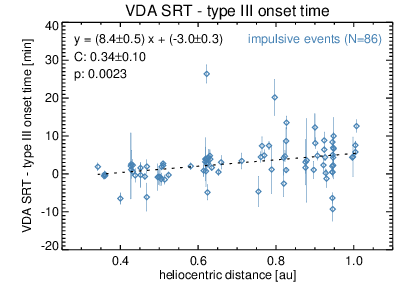}
 \caption{Time difference between SRT (determined with VDA) and solar events plotted versus heliocentric distance. The solar reference times used correspond to the main STIX HXR peak (left) and the RPW type III onset (right). The top row shows gradual events, the bottom row impulsive events. Error bars indicate the one-sigma VDA fit uncertainties. Also shown is the number of events, a linear fit (dotted line) with fit parameters and uncertainties, the Pearson correlation coefficients $C$, and the p-values $p$.}
     \label{fig:corr_vda_dist_compo}
\end{figure*}

\subsection{Timing}

We also addressed the timing of SEE injection relative to the remote-sensing observations of energetic electrons at the Sun, specifically the nonthermal HXR peaks observed by STIX and the type III radio bursts recorded by RPW. Figure~\ref{fig:histo_stix_tdiff}a shows histograms of the time difference between the TSA SRT and the main STIX peak time. The vast majority of events show time differences between -10 min (i.e. injection before main HXR peak) and 40 min, with a median of 6.4 min. Nine events show delays of longer than an hour, the most extreme case being 13 hours. Again, the distributions for impulsive and gradual events are shown as blue and red histograms, respectively. The distribution is narrower for impulsive events, with a median time difference of 5.4 min, while gradual events show a flatter distribution with a median of 14.4 min. 

We note that there are four extreme outliers with TSA SRT delays between 7 and 13 hours. In three of them, the flare association was performed using data from mutiple spacecraft (see Sect.~\ref{sec:epdtiming}). All these events have gradual composition and are characterised by long rise times, ranging from 7.5 to 20 hours. The STIX flare positions show longitude separations (absolute values) from the connectivity footpoint of 39°--161°, with a median of 114°. For comparison, the corresponding median for all events with STIX source locations is 17°. All these characteristics are consistent with SEEs that were accelerated at a CME-driven shock to which Solar Orbiter was poorly connected.

Figure~\ref{fig:histo_stix_tdiff}b  shows the SRT difference distributions for TSA with respect to the type III burst onsets. The results are quite similar to what was found for the main HXR peaks, if anything, the distributions are narrower than for the HXR peaks. Consequently, the timing of flares and radio bursts is quite consistent. The median time difference (flare peak minus type III onset, not shown here) is 0.8 min, and in two thirds of the events, the time difference is less than 5 min. Our TSA SRT delays with respect to type~III onsets are generally consistent with the results of \cite{Cane2003b} for  SEE events observed at 1~au; however, with a range of 1--23~min, the delays of the latter study had a narrower distribution.

While TSA SRTs could be derived for 246 events, VDA fits were only possible for 107 events. The corresponding SRT differences with respect to the main STIX HXR peaks are shown in Fig.~\ref{fig:histo_stix_tdiff}c. The distributions are narrower than for the TSA times, and there are no delays longer than 34 min. Again the X-ray peaks in the impulsive events are better correlated with the SRTs as compared to the gradual events, with median delays of 2.8 min and 10.6 min, respectively. For comparison, the fit uncertainties for the VDA times range from 0.2 to 17.5 min, with 92\% of cases below 5 min and a median of 1.7 min. Finally, Fig.~\ref{fig:histo_stix_tdiff}d shows the SRT difference distributions for VDA with respect to the type III burst onsets. Again, the results are quite consistent with those for the main HXR peaks. For comparison, \cite{Kouloumvakos2015} obtained broadly similar VDA SRT difference distribution for SEE events observed at 1~au, but shifted to longer delays (with a mean of 12.3~min) as compared to our result.

The smaller SRT differences for VDA appear to be largely due to a selection effect, since the TSA time differences for just the events where a VDA fit could be performed show a distribution that is very similar to the VDA times. We note that when VDA is not possible, it is typically due to faint SEE events, instrumental effects, or complications such as variability in the IP medium or previous events that partially mask the event under analysis. In these cases, a TSA time is reported, but the error is likely not given by the data resolution, but much larger, which explains the greater dispersion of the SRT differences compared to VDA.

Table~\ref{tab:timing} summarises the basic statistical parameters of the distributions shown in Fig.~\ref{fig:histo_stix_tdiff}. In particular, note the significantly larger standard deviations of the SRT differences for the gradual events, which is especially pronounced for the TSA times. Comparing these results with the comprehensive study of impulsive SEEs by \cite{Haggerty2002}, we find somewhat shorter delays, i.e. more of the order of 5~min than of 10~min. Conversely, the standard deviations of our TSA SRT difference distributions for impulsive events is slightly higher than found by \cite{Haggerty2002}. Finally, we note that there are more complex events that show different injection times for low-energy and high-energy electrons \citep{Jebaraj2023}.

\begin{table}[h]
\caption[]{Statistics on SRT time differences (TSA and VDA) with respect to the main HXR peaks and the type III onsets, shown separately for impulsive and gradual events.}
\label{tab:timing}
\centering
\begin{tabular}{lcccc}
\hline
\noalign{\smallskip}
event type & number & median & mean & standard dev. \\
 &  & [min] & [min] & [min] \\
\noalign{\smallskip}
\hline
\noalign{\smallskip}
TSA: &  &  &  &  \\
HXR, impuls. & 208 & 5.4 & 8.1 & 13.0 \\
HXR, gradual & 49 & 14.4 & 44.8 & 125.9 \\
type III, impuls. & 196 & 3.4 & 5.5 & 8.7 \\
type III, gradual & 46 & 8.2 & 41.5 & 130.1 \\
\hline
VDA: &  &  &  &  \\
HXR, impuls. & 85 & 2.8 & 3.7 & 6.8 \\
HXR, gradual & 18 & 10.6 & 11.0 & 10.2 \\
type III, impuls. & 86 & 2.3 & 3.0 & 5.2 \\
type III, gradual & 19 & 4.3 & 6.1 & 7.8 \\
            \noalign{\smallskip}
            \hline
\end{tabular}
\end{table}

With Solar Orbiter, we can go beyond distributions and study the SRT difference as a function of distance from the Sun. In case the common time delays are due to propagation effects in the IP medium (as opposed to delayed injection) we would expect to see a correlation of the time delays with distance, provided that the effects act beyond 0.3 au. The TSA times (not shown here) show no obvious correlation with heliocentric distance. However, it should be taken into account that the variability in the conditions of the IP medium and deviations with respect to the ideal Parker spiral length may dominate the observations, masking radial dependences. We thus focus on the subset of events for which we could perform VDA fits. According to \cite{Laitinen2015}, the delays obtained with VDA are the combined result of IP scattering effects and energy-dependent backgrounds affecting the onset determination; therefore, we can expect that possible radial trends in the VDA delays represent, at least in part, how scattering effects accumulate as radial distance increases.

\begin{table*}[]
\caption[]{Correlation of SRT time differences with heliocentric distance for impulsive events, filtered for different additional parameters.}
\label{tab:correls}
\centering
\begin{tabular}{l|ccc|ccc|}
\hline
\hline
\noalign{\smallskip}
filtering &   & STIX main HXR peak & & & RPW type III onset & \\
criteria &  no. & $C$ & $p$ & no. & $C$ & $p$ \\
\noalign{\smallskip}
\hline
\noalign{\smallskip}
all impulsive	&	85	&	$0.30 \pm 0.09$	&	0.008	&	86	&	$0.34 \pm 0.10$	&	0.0023	\\
\hline
single peak/burst 	&	49	&	$0.29 \pm 0.13$	&	0.065	& 29	&	$0.30 \pm 0.16$	&	0.13	\\
multiple peaks/bursts	&	36	&	$0.29 \pm 0.15$	&	0.092		&	57	&	$0.37 \pm 0.13$	&	0.01 \\
\hline
high anisotropy	&	45	&	$0.17 \pm 0.13$	&	0.31	&	44	&	$0.44 \pm 0.13$	&	0.004	\\
medium anisotropy	&	21	&	$0.48 \pm 0.20$	&	0.03	&	22	&	$0.41 \pm 0.26$	& 0.062	\\
\hline
well-connected	&	47	&	$0.23 \pm 0.14$	&	0.15	&	42	&	$0.19 \pm 0.15$	&	0.24	\\
poorly connected	&	19	&	$0.42 \pm 0.16$	&	0.088	&	17	&	$0.57 \pm 0.20$	&	0.039	\\
\hline
ambient SW	&	42	&	$0.52 \pm 0.10$	&	0.0005	&	44	&	$0.46 \pm 0.14$	&	0.002	\\
IP structure	&	15	&	$0.27 \pm 0.17$	&	0.39	&	13	&	$0.46 \pm 0.16$	&	0.13	\\
\noalign{\smallskip}
\hline
\end{tabular}
\tablefoot{
The time differences refer to VDA SRTs with respect to the main STIX HXR peak and the RPW type III onset. Shown are the numbers of events, the Pearson correlation coefficients $C$, and the p-values $p$.}
\end{table*}

\begin{figure*}[]
\centering
  \includegraphics[width=.48\linewidth]{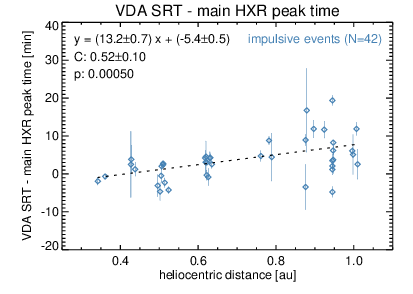}
  \includegraphics[width=.48\linewidth]{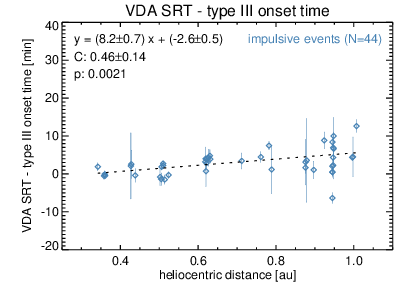}
 \caption{As in Fig.~\ref{fig:corr_vda_dist_compo}, but showing the VDA SRT time difference with respect to the main STIX HXR peak (left) and the RPW type III onset (right) as a function of  heliocentric distance only for impulsive events that were observed in the ambient solar wind, i.e. without the presence of IP structures.}
     \label{fig:corr_vda_dist_compo_ambient}
\end{figure*}

Figure~\ref{fig:corr_vda_dist_compo} shows scatter plots of the VDA SRT differences with respect to the main HXR peak (left panels) and type III burst onset (right panels). We split the events again into gradual (top panels) and impulsive ones (bottom panels). Also shown are linear fits to the data (dotted lines, with fit parameters indicated at the top left of each plot), the Pearson correlation coefficient $C$ (including the uncertainties on $C$ based on a bootstrapping approach), and the p-value $p$. While gradual events do not show any correlation between SRT differences and heliocentric distance, a weak correlation is found in impulsive events, with $C=0.30 \pm 0.09$ and  $0.34 \pm 0.1$ for the main HXR peaks and type III bursts, respectively. There is a trend of the SRT delays increasing from around zero at 0.3 au to $\approx$10 min at 1 au. This is qualitatively consistent with the recent results of \cite{Mitchell2025} who used data from Parker Solar Probe to show that the difference between electron release times (using TSA) and type III burst onset times tend to increase between 0.1 and 0.8~au. The increase in the SRT difference with distance also accounts for the shorter average SRT delays and slightly broader SRT difference distribution as compared to \cite{Haggerty2002} who used observations only at 1~au.

\begin{figure*}[]
\centering
  \includegraphics[width=.48\linewidth]{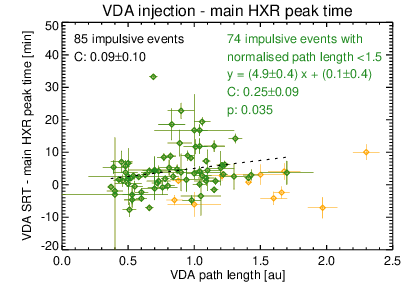}
  \includegraphics[width=.48\linewidth]{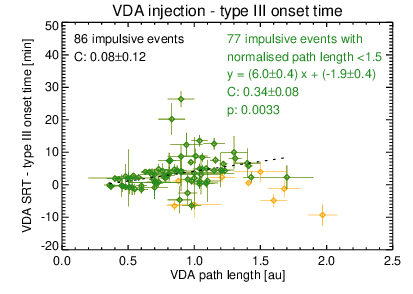}
 \caption{VDA SRT time difference with respect to the main STIX HXR peak (left) and the RPW type III onset (right) as a function of the VDA path length. The green diamonds correspond to the events with normalised path lengths shorter than 1.5 times the nominal Parker spiral lengths, while the rest of the sample is shown with orange diamonds. The plots also indicate the number of events and the correlation coefficients for all events in black, and  for the well-connected events in green. For the latter events, a linear fit, the fit parameters, and the p-value is shown as well. Error bars indicate the one-sigma VDA fit uncertainties for the path length and the SRT.}
     \label{fig:corr_srt_path}
\end{figure*}

Tentative evidence of an influence of distance on the SRT delays was only found when the events were filtered according to their composition. Still, the correlations seen in the impulsive events are weak, and there are several outliers (cf. lower panels in Fig.~\ref{fig:corr_vda_dist_compo}). However, we can now proceed and filter the impulsive events for additional parameters in order to study which of them influence the SRT differences. Table~\ref{tab:correls} compares the number of events and correlation coefficients (only for impulsive events, VDA SRT differences with respect to the main STIX peak and the RPW type III onset) for a range of different classes of events:
\begin{itemize}
    \item Single versus multiple HXR peaks or type III bursts. Here, we checked whether a misidentification of the reference time in events with multiple HXR peaks or type III burst could account for outliers and a weak correlation. For STIX we used the quality rating which distinguishes between events with either a single or a clearly dominating peak and events with multiple peaks of comparable magnitude. We find that misidentification in cases of multiple peaks or bursts is not a major issue. 
    \item Large versus medium anisotropy. Here, the results are more ambiguous. While the correlation is larger for the events of medium anisotropy when the HXR peaks are considered, no difference is found for the type III bursts.
    \item Well-connected versus poorly connected events. This classification is based on a longitude difference between the STIX source and the footpoint of magnetic connectivity of below and above 20$^\circ$, respectively (see Sect.~\ref{sec:conn}). We find significantly higher correlation factors for poorly connected events ($0.42 \pm 0.16$ for the main HXR peaks and $0.57 \pm 0.2$ for  type III bursts) as compared to well-connected ones. This suggests that the progressively increasing delays are at least partly caused by the time it takes for the SEEs to diffuse across magnetic field lines to reach the spacecraft. However, note that we were unable to find a direct correlation between the SRT difference and the longitude difference, as has also been shown by \cite{Krucker1999} and \cite{Mitchell2025}. Note that in contrast to this, \cite{Richardson2014} found that high-energy (0.7–4 MeV) SEE events  generally show shorter delays if the associated flare is located close to the spacecraft foootpoint.
    \item Ambient solar wind versus IP structures. We consider here only events for which the measurements of the IP conditions have a high confidence level; hence, the lower event numbers (see Sect.~\ref{sec:res_context}). The timing of HXR bursts shows a significantly higher correlation in events observed in the quiet solar wind as opposed to IP structures. For type III bursts there is no influence on the correlation coefficient, but the statistical significance of the linear fit is significantly improved when considering the events observed in the quiet solar wind. These results are consistent with magnetic structures that modify the SEE transport, and thus with the notion that the delays are caused by the propagation in the interplanetary medium \citep{Cane2003b}.
\end{itemize}

By filtering impulsive events according to various criteria, we demonstrate that the correlation between SRT difference and heliocentric distance can be moderately strong. An example of these improved correlations is shown in Fig.~\ref{fig:corr_vda_dist_compo_ambient} for impulsive events observed in ambient solar wind conditions. All cases show an increase in SRT delays with heliocentric distance. The slope of this increase varies slightly with filtering criterion but is generally of the order of 10 min per au, which corresponds to zero time difference close to the Sun and 10 min delay at 1 au. We thus show that transport effects that accumulate with distance are at least partially responsible for the observed SRT delays. We also find evidence of the modification of SEE transport by IP structures.

One notable aspect is that the correlations in SRT delays we find tend to be higher when type III bursts are considered, as opposed to the main HXR peaks. This is consistent with the notion that type III bursts are a more direct signature of electron injection into IP space. Indeed in 60 events we find that there is an additional HXR peak closer to the type III onset than the main peak. When we repeat our analysis for the HXR peak times that are most closely associated with the radio bursts (not shown here), we retrieve correlations on levels comparable to the type III bursts or event higher (e.g. $C=0.61 \pm 0.16$ for events in ambient solar wind).

When we consider all impulsive events, we do not find correlations between SRT difference and path length, as shown in Fig.~\ref{fig:corr_srt_path}. This is surprising since the path lengths are proportional to heliocentric distance. The distribution of normalised path lengths (see Fig.~\ref{fig:path}a) shows that events with normalised path lengths $>1.5$ represent outliers. When these events are omitted (highlighted in orange in Fig.~\ref{fig:corr_srt_path}), we obtain correlations of the same order as for the SRT differences as a function of heliocentric distance (see Figs.~\ref{fig:corr_vda_dist_compo}c and ~\ref{fig:corr_vda_dist_compo}d). It appears that the IP structures which are associated with long normalised paths (see Sect.~\ref{sec:paths}) obscure the systematic trends we see in quiet IP conditions, just as we found above when we filtered the events for ambient solar wind conditions. Note that the cases with longer normalised paths mostly show either small or even negative SRT differences, which implies that despite the longer paths travelled the SEEs do not experience the same level of propagation effects as in the cases with more Parker-like configurations. A possible explanation for this is that a large fraction of SEEs with long path lengths are propagating within ICME magnetic flux ropes which are characterised by smooth, low-variance, magnetic fields. The level of magnetic fluctuations in ICME magnetic flux ropes is generally lower than in the typical solar wind \citep{Dasso2005}, and exceptionally long mean free paths (i.e. very low scattering conditions) have been reported for some solar energetic particle events propagating inside these structures \citep{Torsti2004}.

\section{Conclusions}
\label{sec:conclusion}

Through a joint effort of eight of Solar Orbiter's instrument teams (EPD, STIX, EUI, RPW, Metis, SoloHI, SWA, and MAG), we have compiled CoSEE-Cat. As of now, CoSEE-Cat contains the basic parameters of all of the 303 selected SEE events observed by EPD until the end of 2022, as well as information on associated solar events (flares, eruptive phenomena, CMEs, and radio bursts).

In this paper, we present a first statistical analysis of this comprehensive dataset. We studied the distributions of and correlations between various parameters. What we see in nearly all respects is a pronounced difference between events of impulsive and gradual ion composition. Impulsive SEE events tend to have shorter rise times and are more anisotropic. Nearly all SEE events are associated with X-ray and EUV flares and/or eruptive phenomena. The positions of these flares are consistent with the footpoints of the magnetic field lines connecting to the spacecraft in the impulsive events, which indicates that the electrons are accelerated at localised regions with spatial scales of flares or ARs. In contrast, sources associated with gradual events do not show any association with the connecting field lines, which implies that the acceleration region has to be extended or that injection occurs in connection with large magnetic structures. The time difference between the inferred electron release time at the Sun and the nonthermal HXR flare peak or the type III radio burst onset is significantly smaller for impulsive events. Gradual events show more substantial delays of the SRT, supporting the notion that in these events the electrons are not directly injected from the flare site.

Although several of these characteristics were found by previous studies, they are very clearly seen in our sample, most probably due to the very homogeneous dataset provided by state-of-the-art instruments on a single platform, combined with the application of refined analysis methods and data-driven modelling of the magnetic connectivity. We conclude that in impulsive SEE events acceleration in flares and/or small-scale eruptive phenomena like jets is strongly favoured. For gradual SEE events, a more plausible scenario is acceleration at CME-driven shocks, which is further supported by the higher association of gradual events with CMEs. However, more detailed studies of the characteristics of the associated CMEs and type~II radio bursts will be required to conclusively distinguish this model from alternatives such as magnetic reconnection between an erupting flux rope and the ambient open field. 

One of the main unsolved questions of SEEs is the nature of the observed SRT delays: whether they are due to an actually delayed injection at the Sun, or are just apparent delays caused by transport effects. We addressed this issue by investigating the dependency of the release time difference on heliocentric distance. For impulsive events, we find a trend of increasing SRT delay at larger distances. By filtering these events for different parameters, we demonstrate that the trend is more clearly seen under quiet IP conditions (i.e. ambient solar wind, Parker-like IMF). This also highlights the strong influence of the structure of the inner heliosphere on electron propagation. We also find that the trend is more clearly seen for events that are  poorly magnetically connected to Solar Orbiter. Thus, the observed delays could be at least partly explained by the time it takes for the SEEs to diffuse across magnetic field lines to reach the spacecraft.

An important next step in our ongoing investigation will be the spectral analysis of EPD events and STIX flares, which will allow us, for the first time, to compare the spectral parameters of SEEs and flare electrons as functions of heliocentric distance. We shall also take a closer look at the kinematics of associated CMEs using EUI, Metis, SoloHI, and potentially multi-spacecraft observations. More precise timing and starting frequencies of type III bursts can be provided by ground-based metric radio data (including radioheliographic observations), and a search of both ground-based and space-based radio data for type II bursts will better constrain coronal and IP shocks. Finally, we stress that CoSEE-Cat is an ideal starting point for further studies of SEEs using multi-spacecraft in situ observations, particularly to study the angular extent and radial evolution of energetic electron beams.

\section*{Data availability}
CoSEE-Cat is a living catalogue, and as the Solar Orbiter mission progresses, there will be new data releases covering the more recent events (a total of 650 events have been identified until the end of 2024). The catalogue can be accessed online through the  CoSEE-Cat website (\url{https://coseecat.aip.de/}). This provides a web interface with filtering options and access to plots and movies for the individual events, as highlighted in Appendix~\ref{sec:online}.

\begin{acknowledgements}
      Solar Orbiter is a mission of international cooperation between ESA and NASA, operated by ESA. The STIX instrument is an international collaboration between Switzerland, Poland, France, Czech Republic, Germany, Austria, Ireland, and Italy. The EPD/Suprathermal Ion Spectrograph (SIS) is a European facility instrument funded by ESA under contract number SOL.ASTR.CON.00004.  Solar Orbiter post-launch work at JHU/APL is supported by NASA contract NNN06AA01C and at CAU by the German Federal Ministry for Economic Affairs and Energy and the German Space Agency (Deutsches Zentrum für Luft- und Raumfahrt, e.V., (DLR)), grant number 50OT2002. The UAH team acknowledges the financial support by the Spanish Ministerio de Ciencia, Innovaci\'on y Universidades under Project PID2019-104863RB-I00/AEI/10.13039/501100011033 and Project PID2023-150952OB-I00 funded by MICIU/AEI/10.13039/501100011033 and by FEDER, UE. The RPW instrument was built under the responsibility of the French space agency CNES and is an international collaboration between France, Austria, Czech Republic, Germany, Sweden, and USA. The EUI instrument was built by CSL, IAS, MPS, MSSL/UCL, PMOD/WRC, ROB, LCF/IO with funding from the Belgian Federal Science Policy Office (BELPSO); the Centre National d’Etudes Spatiales (CNES); the UK Space Agency (UKSA); the Bundesministerium f\"ur Wirtschaft und Energie (BMWi) through the Deutsches Zentrum f\"ur Luft- und Raumfahrt (DLR); and the Swiss Space Office (SSO). Metis was built and operated with funding from the Italian Space Agency (ASI), under contracts to the National Institute of Astrophysics (INAF) and industrial partners. Metis was built with hardware contributions from Germany (Bundesministerium f\"ur Wirtschaft und Energie through DLR), from the Czech Republic (PRODEX) and from ESA. The Solar Orbiter Heliospheric Imager (SoloHI) instrument was designed, built, and is now operated by the US Naval Research Laboratory with the support of the NASA Heliophysics Division, Solar Orbiter Collaboration Office under DPR NNG09EK11I. Solar Wind Analyser (SWA) data are derived from scientific sensors which have been designed and created, and are operated under funding provided in numerous contracts from the UK Space Agency (UKSA), the UK Science and Technology Facilities Council (STFC), the Agenzia Spaziale Italiana (ASI), the Centre National d’Etudes Spatiales (CNES, France), the Centre National de la Recherche Scientifique (CNRS, France), the Czech contribution to the ESA PRODEX programme and NASA. Solar Orbiter SWA work at UCL/MSSL is currently funded under UKSA/STFC grants ST/X002152/1 and ST/W001004/1. Solar Orbiter magnetometer operations are funded by the UK Space Agency (grant ST/X002098/1).  Solar Orbiter EUI work at UCL/MSSL is currently funded under UKSA grant ST/X002012/1.  

      The AIP team was supported by the German Space Agency (DLR), grant numbers \mbox{50 OT 1904} and \mbox{50 OT 2304}. A.W. and J.M. also acknowledge funding by the European Union’s Horizon Europe research and innovation programme under grant agreement No. 101134999 (SOLER).

      N.D. is grateful for support by the Academy of Finland (SHOCKSEE, grant No.\ 346902). 

      D.P. acknowledges the support by the National Natural Science Foundation of China (Grant Nos. 42188101 and 42130204).

      M.K. is acknowledging funding from CNES for the Solar Orbiter/RPW project. N.V and D.P-L acknowledge support from CNES for the Solar Orbiter/STIX project.
      
The ROB team thank the Belgian Federal Science Policy Office (BELSPO) for the provision of financial support in the framework of the PRODEX Programme of the European Space Agency (ESA) under contract numbers 4000112292, 4000134088, 4000106864, 4000134474, and 4000136424.
       L.R.-G.\ and S. M.\ acknowledge support through the European Space Agency (ESA) research fellowship programme.
      
      Research was sponsored by the NASA Goddard Space Flight Center through a contract with ORAU. The views and conclusions contained in this document are those of the authors and should not be interpreted as representing the official policies, either expressed or implied, of the NASA Goddard Space Flight Center or the U.S. Government. The U.S.Government is authorised to reproduce and distribute reprints for Government purposes notwithstanding any copyright notation herein. F.C. acknowledges the financial support by an appointment to the NASA Postdoctoral Program at the NASA Goddard Space Flight Center, administered by ORAU through a contract with NASA, and the support of the Solar Orbiter mission.

    H.R. acknowledges the support from the STFC grant ST/W001004/1.
           
    F.E. acknowledges support by the German Science Foundation (DFG) SFB grant 1491 and by the International Space Science Institute (ISSI) in Bern, through ISSI International Team project 24-608 (Energetic Particle Transport in Space Plasma Turbulence).

 This work was supported by the long-term programme of support of the Ukrainian research teams at the Polish Academy of Sciences carried out in collaboration with the U.S. National Academy of Sciences with the financial support of external partners.
 
       We thank K.-L.\,Klein for the insightful comments and suggestions provided.

\end{acknowledgements}

%
%

\bibliographystyle{aa} 
\bibliography{references}

\begin{appendix}
\section{Content of CoSEE-Cat first data release}
\label{sec:contents}

In the following, we list the parameters contained in the columns of CoSEE-Cat first data release. For the determination of these parameters, see Sect.~\ref{sec:catalog}. Note that all times reported in the catalogue refer to their measurement at Solar Orbiter. Inferred SRTs at the Sun have been corrected for light-travel time in order to conform to this frame of reference.

\begin{itemize}
\item {\em event\_id}: unique identifier of each SEE event in the catalogue, encodes the EPD onset time (e.g. 2011170942 for the event of 2020 Nov 17 with the EPD onset at 09:42 UT).
\item {\em event\_date}: refers to the date of the EPD onset.
\item {\em solo\_dist}: Solar Orbiter heliocentric distance (au).
\item {\em solo\_lon}: Solar Orbiter longitude\footnote{\label{coord_solo}In Heliocentric Earth Equatorial (HEEQ) coordinates} (°).
\item {\em solo\_lat}: Solar Orbiter latitude\footref{coord_solo} (°).

\bigskip
The following columns provide information on the electron events as measured by EPD:

\item {\em epd\_tonset}: time of energetic electron onset at Solar Orbiter (UT).
\item {\em epd\_delta\_t}: time resolution used to estimate electron onset time (s).
\item {\em epd\_tpeak}: time of maximum electron intensity (UT).
\item {\em epd\_ipeak}: maximum electron intensity ($\mathrm{s^{-1}~cm^{-2}~sr^{-1}~MeV^{-1}}$); pre-event background has not been subtracted.
\item {\em epd\_epeak}: energy at which the peak intensity was measured (keV). As a standard, 44~keV was adopted, but in weaker events lower energies had to be used.
\item {\em tsa\_itime}: extrapolated electron SRT based on time-shift analysis (UT).
\item {\em vda\_itime}: extrapolated electron SRT based on velocity dispersion analysis (UT).
\item {\em vda\_time\_unc}: timing uncertainty given by standard deviation of VDA fit (minutes).
\item {\em vda\_lpath}: path length given by VDA fit (au).
\item {\em vda\_path\_unc}: path length uncertainty given by standard deviation of VDA fit (au).
\item {\em nominal\_path}: nominal path length along the Parker spiral (au).
\item {\em epd\_aniso}: degree of anisotropy (high, medium, small).
\item {\em epd\_compo}: composition of associated energetic ions (impulsive, gradual, intermediate).
\item {\em epd\_series}: is this event part of a series? (y - yes, n - no).
\item {\em epd\_dispersive}: does this event have a dispersive onset? (y - yes, n - no).

\bigskip
The following columns give the parameters of the associated solar flares based on X-ray observations:

\smallskip
\item {\em goes\_class}: GOES flare class\footnote{Only provided for events that are visible from Earth and flares are listed in the solar event reports issued by the Space Weather Prediction Center.}.
\item {\em goes\_estim}: estimated GOES flare class, based on the STIX counts in the 4-10 keV energy range.
\item {\em stix\_status}: STIX instrument status: 1 (nominal) or 0 (not observing).
\item {\em stix\_tpeak}: main STIX peak time (UT).
\item {\em stix\_emax}: maximal energy at which the main STIX peak was detected (keV).
\item {\em stix\_tpeak\_epd}: time of STIX peak closest to inferred SRT (UT).
\item {\em stix\_tpeak\_rpw}: time of STIX peak closest to type III radio burst onset time (UT).
\item {\em stix\_hpc\_x}: STIX source X-coordinate\footnote{\label{coord_stix_hpc}In Helioprojective Cartesian coordinates} (asec).
\item {\em stix\_hpc\_y}: STIX source Y-coordinate\footref{coord_stix_hpc} (asec).
\item {\em stix\_hgs\_lat}: STIX source latitude\footnote{\label{coord_stix_hgc}In Heliographic coordinates} (°).
\item {\em stix\_hgs\_lon}: STIX source Stonyhurst longitude\footref{coord_stix_hgc} (°).
\item {\em stix\_carr\_lon}: STIX source Carrington longitude \footref{coord_stix_hgc} (°).
\item {\em stix\_ar\_num}: NOAA active region number associated with STIX source.
\item {\em stix\_epd\_qual}: confidence level of STIX-EPD association, from 1 (high) to 3 (low).
\bigskip
\\
We continue with the columns that provide parameters on the associated EUI flares and/or eruptions:
\item {\em eui\_status}: EUI instrument status: 1 (nominal), 0 (not observing), or "cor" when data was acquired in coronographic mode.
\item {\em eui\_hpc\_x1}: EUI primary source X-coordinate\footref{coord_stix_hpc} (asec).
\item {\em eui\_hpc\_y1}: EUI primary source Y-coordinate\footref{coord_stix_hpc} (asec).
\item {\em eui\_hgs\_lat1}: EUI primary source latitude\footref{coord_stix_hgc} (°).
\item {\em eui\_hgs\_lon1}: EUI primary source Stonyhurst longitude\footref{coord_stix_hgc} (°).
\item {\em eui\_carr\_lon1}: EUI primary source Carrington longitude \footref{coord_stix_hgc} (°).
\item {\em eui\_ar\_num1}: NOAA active region number associated with EUI primary source.
\item {\em eui\_type1}: eruption type associated with EUI primary source.
\item {\em eui\_hpc\_x2}: EUI 2nd source X-coordinate\footref{coord_stix_hpc} (asec).
\item {\em eui\_hpc\_y2}: EUI 2nd source Y-coordinate\footref{coord_stix_hpc} (asec).
\item {\em eui\_hgs\_lat2}: EUI 2nd source latitude\footref{coord_stix_hgc} (°).
\item {\em eui\_hgs\_lon2}: EUI 2nd source Stonyhurst longitude\footref{coord_stix_hgc} (°).
\item {\em eui\_carr\_lon2}: EUI 2nd source Carrington longitude \footref{coord_stix_hgc} (°).
\item {\em eui\_ar\_num2}: NOAA active region number associated with EUI 2nd source.
\item {\em eui\_type2}: eruption type associated with EUI 2nd source.
\item {\em eui\_hpc\_x3}: EUI 3rd source X-coordinate\footref{coord_stix_hpc} (asec).
\item {\em eui\_hpc\_y3}: EUI 3rd source Y-coordinate\footref{coord_stix_hpc} (asec).
\item {\em eui\_hgs\_lat3}: EUI 3rd source latitude\footref{coord_stix_hgc} (°).
\item {\em eui\_hgs\_lon3}: EUI 3rd source Stonyhurst longitude\footref{coord_stix_hgc} (°).
\item {\em eui\_carr\_lon3}: EUI 3rd source Carrington longitude \footref{coord_stix_hgc} (°).
\item {\em eui\_ar\_num3}: NOAA active region number associated with EUI 3rd source.
\item {\em eui\_type3}: eruption type associated with EUI 3rd source.
\bigskip
\\
The next columns give information on the associated radio bursts:
\item {\em rpw\_status}: RPW instrument status: 1 (nominal) or 0 (not observing).
\item {\em rpw\_t3\_time}: type III burst starting time (UT).
\item {\em rpw\_t3\_freq}: frequency at which starting time was measured (MHz).
\item {\em rpw\_t3\_num}: number of individual bursts (single, multiple, or storm).
\item {\em rpw\_t2}: presence of type II burst (yes, no).
\bigskip
\\
Associated Metis CMEs are characterised in the next columns:
\item {\em metis\_status}: Metis instrument status: 1 (nominal) or 0 (not observing).
\item {\em metis\_tstart1}: primary CME start time (UT).
\item {\em metis\_tend1}: primary CME end time (UT).
\item {\em metis\_fov\_min1}: Metis FOV inner edge for primary CME (R$_\sun$).
\item {\em metis\_fov\_max1}: Metis FOV outer edge for primary CME (R$_\sun$).
\item {\em metis\_cpa1}: primary CME central position angle (°).
\item {\em metis\_width1}: primary CME angular width (°).
\item {\em metis\_speed1}: primary CME radial speed (km\,s$^{-1}$).
\item {\em metis\_tlaunch1}: primary CME launch time extrapolated to one R$_\sun$ (UT).
\item {\em metis\_tstart2}: second CME start time (UT).
\item {\em metis\_tend2}: second CME end time (UT).
\item {\em metis\_fov\_min2}: Metis FOV inner edge for second CME (R$_\sun$).
\item {\em metis\_fov\_max2}: Metis FOV outer edge for second CME (R$_\sun$).
\item {\em metis\_cpa2}: second CME central position angle (°).
\item {\em metis\_width2}: second CME angular width (°).
\item {\em metis\_speed2}: second CME radial speed (km\,s$^{-1}$).
\item {\em metis\_tlaunch2}: second CME launch time extrapolated to one R$_\sun$ (UT).
\bigskip
\\
The next columns indicate whether a CME was observed by SoloHI:
\item {\em solohi\_status}: SoloHI instrument status: 1 (nominal) or 0 (not observing).
\item {\em solohi\_tstart}: CME Starting time in SoloHI FOV (UT).
\bigskip
\\
The conditions of the interplanetary medium in the context of the events is given in the
following columns. Time differences are measured between the SEE onset and the encounter with each structure (negative: the structure crossed Solar Orbiter before the SEE onset; positive: the structure crossed after the SEE onset; zero: the onset occurs during the crossing of the structure):
\item {\em sw\_speed}: solar wind speed (km\,s$^{-1}$): As measured by SWA/PAS. If no measurements are available, 400~km\,s$^{-1}$ have been adopted for the derivation of TSA times and magnetic connectivity.
\item {\em shock\_delay}: shock passage time difference (hrs).
\item {\em icme\_sheath\_delay}: ICME sheath passage time difference (hrs).
\item {\em icme\_mo\_delay}: ICME magnetic obstacle passage time difference (hrs).
\item {\em post\_icme\_delay}: post-ICME passage time difference (hrs).
\item {\em sir\_comp\_delay}: SIR compression region passage time difference (hrs).
\item {\em sir\_si\_delay}: SIR stream interface passage time difference (hrs).
\item {\em sir\_rar\_delay}: SIR rarefaction region passage time difference (hrs).
\item {\em hcs\_delay}: HCS passage time difference (hrs).
\item {\em ss\_fr\_delay}: SS FR passage time difference (hrs).
\item {\em sw\_condition}: ambient solar wind conditions (f - fast, s - slow, u - unknown).
\item {\em mag\_polarity}: polarity of magnetic field (p - positive, n - negative, a - ambiguous).
\item {\em ip\_complex}: indicates whether the  interplanetary conditions are complex (e.g. multiple structures or undefined IMF);  yes / no 
\item {\em ip\_confidence}: confidence level: from 1 (certain) to 3 (not sure).
\bigskip
\\
Finally, the last columns of the catalogue provide information on the magnetic connectivity:
\item {\em conn\_carr\_lon}: Carrington longitude of connectivity footpoint\footref{coord_stix_hgc} (°).
\item {\em conn\_carr\_lon\_unc}: connectivity longitude uncertainty (°).
\item {\em conn\_hgs\_lat}: latitude of connectivity footpoint\footref{coord_stix_hgc} (°).
\item {\em conn\_hgs\_lat\_unc}: connectivity latitude uncertainty (°).
\item {\em conn\_conf}: connectivity confidence level: from 1 (high) to 4 (low).

\end{itemize}

\section{The CoSEE-Cat website}
\label{sec:online}

The first data release of CoSEE-Cat is accessible online at \url{https://coseecat.aip.de/}\footnote{\url{https://doi.org/10.17876/coseecat/dr.1}}. This webpage (see Fig.~\ref{fig:online_banner}) gives access to all the parameters described in Appendix~\ref{sec:contents} for the 303 events detected until the end of 2022. In addition, this data release provides the following plots for each individual event:
\begin{itemize}
\item EPD-RPW-STIX overview plot (cf. Fig.~\ref{fig:epd_stix_rpw_example})
\item EPD pitch-angle coverage and anisotropy plot (cf. Fig.~\ref{fig:anisotropy_example})
\item STIX-EUI source location and connectivity plot (cf. Fig.~\ref{fig:stix_eui_example})
\item RPW-STIX plot (cf. Fig.~\ref{fig:rpw_stix_example})
\item  interplanetary context plot (cf. Fig.~\ref{fig:ip_context_example})
\end{itemize}

We also provide links to the following movies:
\begin{itemize}
\item EUI/FSI full-disc daily movies for all events where data is available (both at 174~\AA\ and 304~\AA)
\item Metis running difference movies (visual channel) for all detected CMEs 
\item SoloHI running difference movies for CMEs detected with the instrument
\end{itemize}

\begin{figure*}[]
\centering
  \includegraphics[width=.99\linewidth]{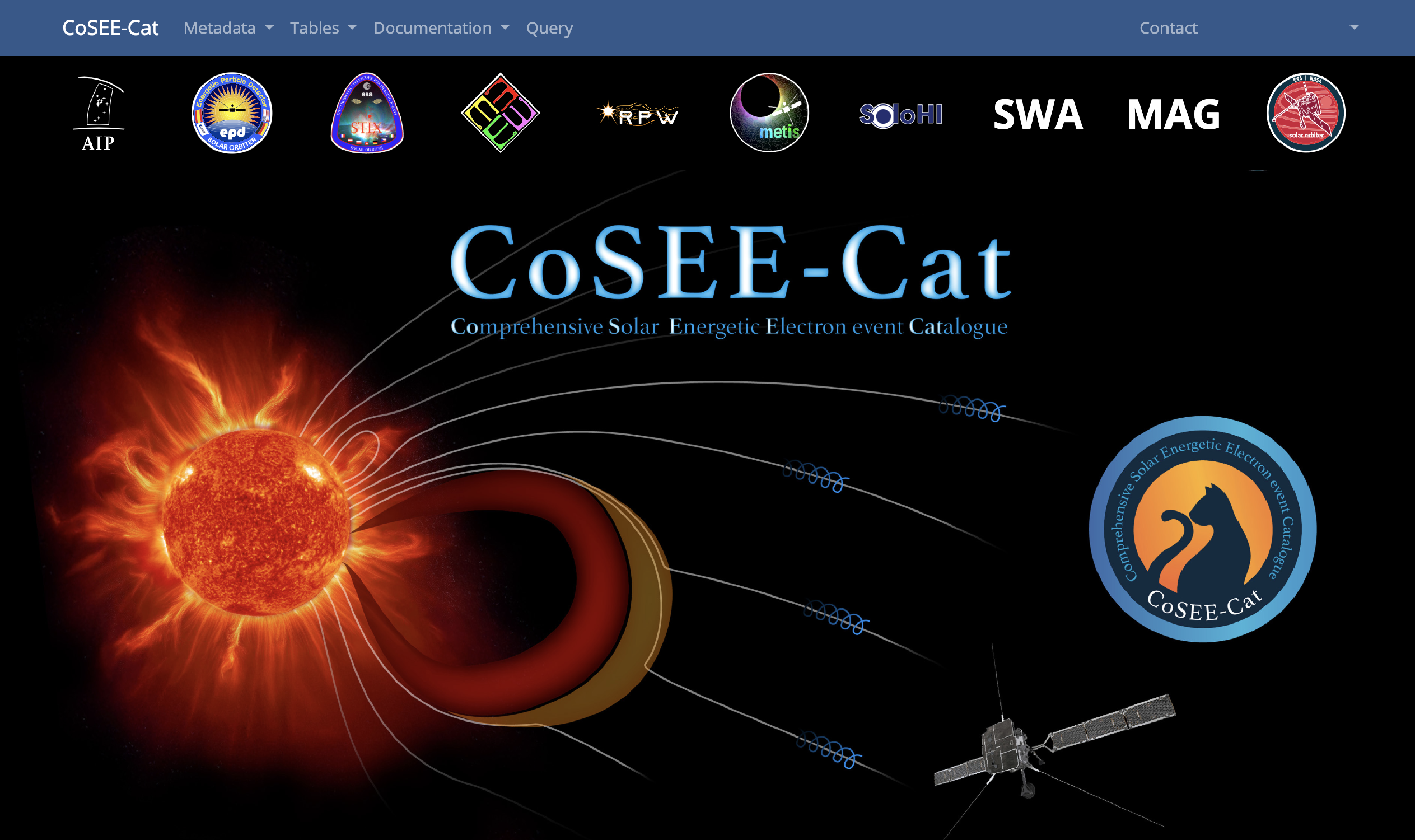}
  \caption{Screenshot of the upper part of the home page for the CoSEE-Cat project.}
     \label{fig:online_banner}
\end{figure*}

The data for each event can be visualised in a tabular form. In addition, for each event, a link is provided to open an event viewer, which displays all the plots available for that event and a sub-set of the most relevant data.
The full content of the data release can be download as files in CSV format. A query interface is also provided, where one can use the full power of the SQL language to search for events that meet some criteria and select the columns to be displayed.
A complete description of the content of the website and its functionalities is available online, in menu ‘Documentation’.

\end{appendix}
  
\end{document}